\documentclass[12pt]{article}
\usepackage{amssymb,epsfig}

\renewcommand{\theequation}{\arabic{section}.\arabic{equation}}

\newcommand{\bra}[1]{\left\langle #1 \right|}
\newcommand{\ket}[1]{\left| #1 \right\rangle}

\newcommand{\beq}[1]{\begin{equation}\label{#1}}
\newcommand{\eeq}{\end{equation}}
\newcommand{\beqar}[1]{
\begin{eqnarray}\label{#1}}
\newcommand{\eeqar}{\end{eqnarray}}

\newcommand{\lash}[1]{\not\! #1 \,}

\newcommand{\nn}{\nonumber}
\newcommand{\dd}{{\rm d}} 
\newcommand{\Gl}[1]{Eq.~(\ref{#1})}
\newcommand{\Ab}[1]{Fig.~\ref{#1}}
\newcommand{\Ta}[1]{Tab.~(\ref{#1})}
\newcommand{\vecl}[1]{\overleftarrow #1}
\newcommand{\vecr}[1]{\overrightarrow #1}
\newcommand{\Dlr}{\stackrel{\leftrightarrow}{D}}

\newcommand{\pz}{{p\cdot z}}
\newcommand{\D}{{\cal D}}
\newcommand{\PS}{{\cal P}}
\newcommand{\V}{{\cal V}}
\newcommand{\A}{{\cal A}}
\newcommand{\T}{{\cal T}}
\newcommand{\up}{{\uparrow}}
\newcommand{\down}{{\downarrow}}
\newcommand{\al}{\alpha}
\newcommand{\be}{\beta}  
\newcommand{\ep}{\varepsilon}
\newcommand{\ga}{\gamma}
\newcommand{\de}{\delta}

\newcommand{\la}{\lambda}

\newcommand{\si}{\sigma}
\newcommand{\ro}{\varrho}
\newcommand{\Ga}{\Gamma}

\newcommand{\La}{\Lambda}

\begin{document}


\begin{titlepage}

\vspace*{-2cm}
\begin{flushright}
\begin{tabular}{l}
TPR-00-10\\
hep-ph/0007279
\end{tabular}
\end{flushright}

\vskip2.5cm

\begin{center}
{\Large \bf Higher Twist Distribution Amplitudes of the Nucleon in QCD}
\vspace{1cm}
\end{center}
\centerline{\large\sc V.~Braun\footnote{
   \it E-mail: Vladimir.Braun@physik.uni-regensburg.de},
    R.~J.~Fries\footnote{
   \it E-mail: Rainer.Fries@physik.uni-regensburg.de},
   N.~Mahnke\footnote{
   \it E-mail: Nils.Mahnke@physik.uni-regensburg.de},
 and E.~Stein\footnote{
   \it E-mail: Eckart.Stein@physik.uni-regensburg.de},}
\vspace{1 cm}
\centerline{{\em  Institut f\"ur Theoretische Physik , 
Universit\"at Regensburg, D-93040 Regensburg, Germany}} 
\vspace{1cm}
\bigskip
\centerline{\large \em \today}
\bigskip
\vfill
\begin{center}
  {\large\bf Abstract\\[10pt]} \parbox[t]{\textwidth}{ 
  We present the first systematic study of higher-twist light-cone distribution
  amplitudes of the nucleon in QCD. We find that the valence three-quark 
  state is described at small transverse separations by eight   
  independent distribution amplitudes. One of them is leading twist-3, 
  three distributions are twist-4 and twist-5, respectively, 
  and one is twist-6.
  A complete set of distribution amplitudes is constructed, which satisfies 
  equations of motion and constraints that follow from conformal expansion. 
  Nonperturbative input parameters are estimated from QCD sum rules.
                                   }
  \vskip1cm 
{\em Submitted to Nuclear Physics B}\\[1cm]
\end{center}

\vspace*{1cm}

\noindent{\bf PACS} numbers: 
\\

\noindent{\bf Keywords:} QCD, Nucleon, Power Corrections, Distribution Amplitudes
\vspace*{\fill}
\eject
\end{titlepage}


\section{Introduction}
\setcounter{equation}{0}

Hard exclusive processes are coming to the forefront of high energy 
nuclear and particle physics. This is already visible at TJNAF, 
COMPASS, HERMES that go for more and more exclusive  channels,
and it makes the core of the ELFE proposal. One main reason for that 
is the growing understanding that exclusive reactions provide unmatched 
opportunities to study the hadron structure, as demonstrated  by recent 
interest to deeply virtual Compton scattering (DVCS) and hard diffractive 
meson production. All future plans also call for very high luminosity 
and would therefore be perfectly suited for the investigation of 
exclusive and semi-exclusive reactions with and without polarization. 

The classical theoretical framework for the calculation of hard 
exclusive processes in QCD was developed in \cite{exclusive,earlybaryon}, see 
\cite{CZreport,BLreport,SSreport} for a review. 
This approach introduces a concept of hadron distribution amplitudes
as fundamental nonperturbative functions describing the hadron structure 
in rare parton configurations with a minimum number of Fock constituents 
(and at small transverse separations). The distribution amplitudes are 
equally important and to a large extent complementary to conventional 
parton distributions which correspond to one-particle probability
distributions for the parton momentum fraction in an  {\em average}  
configuration. The recent data from CLEO \cite{CLEO} and E791 \cite{Ashery} 
provided, for the first time, quantitative information on the pion 
distribution amplitude.

For a theorist,
the main challenge is to make the QCD description of hard exclusive reactions
fully quantitative.  Although the leading-twist QCD factorization approach
correctly reproduces the power dependence of exclusive amplitudes on the 
large momentum, there exist many indications that soft end-point and 
higher-twist corrections might dominate over the naive QCD prediction over 
the range of accessible values of $Q^2$.  
For mesons, the study of preasymptotic corrections to exclusive reactions
has been pursued intensively and in different directions. In particular,
meson distribution amplitudes of higher twist have been studied in detail 
in \cite{CZreport,Bra90,BallBraun}. 
For baryons, the corresponding task is more complicated and 
received less attention in the past. Soft end-point contributions
to several hard exclusive reactions involving nucleons have been estimated 
using various models and were found to be large \cite{softnucleon}.
To the best of our knowledge, hard higher-twist corrections and nucleon mass 
corrections have never been addressed in the literature.    
In this paper we present the first systematic study of nucleon distribution 
amplitudes of higher twist.

The notion of higher twist contributions to the distribution amplitudes 
comprises a broad spectrum of  effects of different physical origin:
 First, contributions of ``bad'' components in the wave function and 
 in particular of components with ``wrong'' spin projection; second, 
 contributions of transverse motion of quarks (antiquarks) in the 
 leading twist components; third, contributions of higher Fock states 
 with additional gluons and/or quark-antiquark pairs. Finally,
 one can speak of hadron mass effects, similar to target mass corrections
 in deep inelastic scattering.
The relative significance of these corrections depends on the  hadron 
state in question and the particular hard process. There exists 
an important conceptual difference between mesons and baryons. 
For mesons, all effects due to ``bad'' components in the quark-antiquark
wave functions can be rewritten in terms of 
higher Fock state components thanks to the QCD equations of motion.
Since quark-antiquark-gluon matrix elements between the vacuum and 
the meson state are numerically small (see \cite{Bra90,BallBraun} for 
the existing  estimates), contributions of ``bad'' components in 
two-particle wave functions are small 
as well, and to a large extent dominated by meson mass corrections
\cite{Bra90,BallBraun}. For baryons, on the contrary, QCD equations of 
motion are not sufficient to eliminate the higher-twist three-quark states 
in favor of the components with extra gluons, so that the former 
present genuine new degrees of freedom. Moreover, 
vacuum-to-baryon matrix elements of higher-twist three-quark operators 
are large,  comparable to the ones of leading twist.
One may expect, therefore, that higher-twist effects in baryon 
wave functions are dominated by ``bad'' components of three-quark states
rather than extra gluons. A systematic study of such states as well as 
taking into account nucleon mass corrections presents the subject of 
this work. 

We find that a generic three-quark matrix element on the light-cone 
can be parametrized in terms of eight independent distribution amplitudes.
In particular, there exists one  amplitude of leading twist-three    
that is familiar from earlier studies \cite{earlybaryon,Che84}, 
three distribution amplitudes of twist-four and twist-five, respectively,
and one distribution amplitude of twist-six. 
Following the approach of \cite{Bra90,BallBraun}, we further expand all 
distribution amplitudes in contributions of operators with increasing 
conformal spin. The conformal expansion corresponds, physically, 
to the separation of longitudinal and transverse degrees of freedom, and  
the coefficients of this expansion present the relevant nonperturbative 
parameters.
We find that the higher-twist distribution amplitudes with the lowest 
conformal spin (asymptotic wave functions) involve two nonperturbative 
matrix elements that are well known from the QCD sum rule studies of 
the nucleon \cite{Iof81,Chu82,Kol84}.
The leading corrections to the asymptotic wave functions involve three 
more parameters (for all twists) that we estimate in this work.     

The outline of the paper is as follows.
The general classification of twist-three distribution amplitudes is 
worked out in Section 2. One starts with 24 invariant functions 
in a general decomposition of a three quark light-cone operator. 
By isospin symmetry the number of independent amplitudes
is reduced to eight amplitudes that determine the three-quark
proton state completely. A convenient representation of these
independent amplitudes is derived in terms of chiral fields.

Section 3 is devoted to the conformal expansion.
We identify the nonperturbative parameters that  describe the  
higher-twist distribution amplitudes up to next-to-leading order 
in the conformal expansion and work out the relations 
between them, as given by isospin constraints and equations of motion.

In Section 4 we present particular models for the higher-twist
distributions based on QCD sum rules and summarize.

Our work encloses a number of appendices. Appendix A gives the
Fierz transformation rules used to derive the isospin relations.
Appendix B summarizes the determination of nonperturbative 
parameters in the QCD sum rule approach.
Appendix C contains a ``handbook'' of nucleon distribution amplitudes 
where we collect all expressions needed in practical calculations.


\section{General classification}
\subsection{Lorentz structure}

The notion of hadron distribution amplitudes in general refers to 
hadron-to-vacuum matrix elements of nonlocal operators built of quark and 
gluon fields at light-like separations. In this paper we will deal with 
the three-quark matrix element 
\beqar{dreiquark}
\bra{0} \ep^{ijk} u_\al^{i'}(a_1 z)\left[a_1 z, a_0 z \right]_{i',i}
u_\be^{j'}(a_2 z) \left[a_2 z, a_0 z \right]_{j',j}
d_\ga^{k'}(a_3 z) \left[a_3 z, a_0 z \right]_{k',k}
\ket{P(P,\la)}
\eeqar
where $\ket{P(P,\la)}$ denotes the proton state with 
momentum $P$, $P^2 = M^2$ and helicity $\la$. 
$u,d$ are the quark-field operators. The Greek letters $\al,\be,\ga$ 
stand for Dirac indices, the Latin letters $i,j,k$ refer to color.
$z$ is an arbitrary light-like  vector, $z^2=0$, the $a_i$ are real numbers.
The gauge-factors $\left[x,y\right]$ are defined as
\beqar{path}
\left[x,y\right] = {\rm P}\exp\left[ig\int_0^1\! \dd t\, (x-y)_\mu 
A^\mu(t x + (1-t) y)\right]\,
\eeqar
and render the matrix element in (\ref{dreiquark}) gauge-invariant.
To simplify the notation we will not write the gauge-factors explicitely
in what follows but imply that they are always present.

Taking into account Lorentz covariance, spin and parity of the
nucleon, the most general decomposition  of the matrix element in
\Gl{dreiquark} involves 24 invariant functions:  
\beqar{zerl}
\lefteqn{ 4 \bra{0} \ep^{ijk} u_\al^i(a_1 z) u_\be^j(a_2 z) d_\ga^k(a_3 z) 
          \ket{P} =}
\nn \\
&=&  
{\cal S}_1 M C_{\al \be} \left(\ga_5 N\right)_\ga + 
{\cal S}_2 M^2 C_{\al \be} \left(\!\not\!{z} \ga_5 N\right)_\ga + 
{\cal P}_1 M \left(\ga_5 C\right)_{\al \be} N_\ga + 
{\cal P}_2 M^2 \left(\ga_5 C \right)_{\al \be} \left(\!\not\!{z} N\right)_\ga 
\nn \\
&& + 
{\cal V}_1  \left(\!\not\!{P}C \right)_{\al \be} \left(\ga_5 N\right)_\ga + 
\V_2 M \left(\!\not\!{P} C \right)_{\al \be} \left(\!\not\!{z} \ga_5 N\right)_\ga  + 
\V_3 M  \left(\ga_\mu C \right)_{\al \be}\left(\ga^{\mu} \ga_5 N\right)_\ga 
\nn \\ 
&& +
\V_4 M^2 \left(\!\not\!{z}C \right)_{\al \be} \left(\ga_5 N\right)_\ga +
\V_5 M^2 \left(\ga_\mu C \right)_{\al \be} \left(i \si^{\mu\nu} z_\nu \ga_5 
N\right)_\ga 
+ \V_6 M^3 \left(\!\not\!{z} C \right)_{\al \be} \left(\!\not\!{z} \ga_5 N\right)_\ga  
\nn \\ 
&& + 
\A_1  \left(\!\not\!{P}\ga_5 C \right)_{\al \be} N_\ga + 
\A_2 M \left(\!\not\!{P}\ga_5 C \right)_{\al \be} \left(\!\not\!{z} N\right)_\ga  + 
\A_3 M  \left(\ga_\mu \ga_5 C \right)_{\al \be}\left( \ga^{\mu} N\right)_\ga 
\nn \\ 
&& +
\A_4 M^2 \left(\!\not\!{z} \ga_5 C \right)_{\al \be} N_\ga +
\A_5 M^2 \left(\ga_\mu \ga_5 C \right)_{\al \be} \left(i \si^{\mu\nu} z_\nu  
N\right)_\ga 
+ \A_6 M^3 \left(\!\not\!{z} \ga_5 C \right)_{\al \be} \left(\!\not\!{z} N\right)_\ga  
\nn \\
&& +
\T_1 \left(P^\nu i \si_{\mu\nu} C\right)_{\al \be} 
\left(\ga^\mu\ga_5 N\right)_\ga + 
\T_2 M \left(z^\mu P^\nu i \si_{\mu\nu} C\right)_{\al \be} 
\left(\ga_5 N\right)_\ga +
\T_3 M \left(\si_{\mu\nu} C\right)_{\al \be} 
\left(\si^{\mu\nu}\ga_5 N\right)_\ga 
\nn \\
&& 
+ 
\T_4 M \left(P^\nu \si_{\mu\nu} C\right)_{\al \be} 
\left(\si^{\mu\ro} z_\ro \ga_5 N\right)_\ga 
+ 
\T_5 M^2 \left(z^\nu i \si_{\mu\nu} C\right)_{\al \be} 
\left(\ga^\mu\ga_5 N\right)_\ga 
\nn \\
&& +
\T_6 M^2 \left(z^\mu P^\nu i \si_{\mu\nu} C\right)_{\al \be} 
\left(\!\not\!{z} \ga_5 N\right)_\ga  
+
\T_{7} M^2 \left(\si_{\mu\nu} C\right)_{\al \be} 
\left(\si^{\mu\nu} \!\not\!{z} \ga_5 N\right)_\ga
\nn \\
&&+ 
\T_{8} M^3 \left(z^\nu \si_{\mu\nu} C\right)_{\al \be} 
\left(\si^{\mu\ro} z_\ro \ga_5 N\right)_\ga \,,
\eeqar
where $N_\ga$ is the nucleon spinor, $C$ the charge conjugation matrix 
and  $\si_{\mu\nu} = \frac{i}{2} [\ga_\mu,\ga_\nu]$. The factor 4 
on the l.h.s. is introduced for later convenience.
Each of the 24 functions  ${\cal S}_i,\PS_i, \A_i, \V_i, \T_i$ 
depends on the scalar product $P\cdot z$.

The invariant functions in \Gl{zerl} do not have a definite twist yet.
For the twist classification, it is convenient to go over to the 
infinite momentum frame. To this end we introduce the second light-like 
vector 
\beqar{vectors}
p_\mu &=& P_\mu  - \frac{1}{2} z_\mu \frac{M^2}{p\cdot z}\,,~~~~~ p^2=0\,, 
\eeqar
so that $P \to p$ if the nucleon mass can be neglected $M \to 0$.
Assume for a moment that the nucleon moves in the positive 
${\bf e_z}$ direction, then $p^+$ and $z^-$ are the only nonvanishing 
components of $p$ and $z$, respectively. 
The infinite momentum frame can be visualized
as the limit $p^+ \sim Q \to \infty$ with fixed $P\cdot z = p \cdot z\sim 1$ 
where $Q$ is the large scale in the process.
Expanding the matrix element in powers of $1/p^+$ introduces
the power counting in $Q$. In this language twist counts the
suppression in powers of $p^+$. Similarly, 
the nucleon spinor $N_\ga(P,\la)$ has to be decomposed 
in ``large'' and ``small'' components as
\beqar{spinor}
N_\ga(P,\la) = \frac{1}{2 p\cdot z} \left(\!\not\!{p}\! \!\not\!{z} +
\!\not\!{z}\!\!\not\!{p} \right) N_\ga(P,\la)= N^+_\ga(P,\la) + N^-_\ga(P,\la)
\; ,
\eeqar
where we have introduced two projection operators
\beqar{project}
\Lambda^+ = \frac{\!\not\!{p}\! \!\not\!{z}}{2 p\cdot z} \quad ,\quad
\Lambda^- = \frac{\!\not\!{z}\! \!\not\!{p}}{2 p\cdot z}
\eeqar
that project onto the ``plus'' and ``minus'' components of the spinor.
Note the useful relations
\beqar{bwgl}
\lash{p} N(P) = M N^+(P)\,,&\qquad& \lash{z} N(P) = \frac{2 p z}{M} N^-(P)
\eeqar
that follow readily from the Dirac equation $\lash{P} N(P) = M N(P)$.

Using the explicit expressions for $N(P)$ it is easy to see 
that $\La^+N = N^+ \sim \sqrt{p^+}$ while $\La^-N = N^- \sim 1/\sqrt{p^+}$.
To give an example of  how such power counting works, decompose 
the Lorentz structure in front of $\V_1$ in \Gl{zerl} in terms of 
light-cone vectors
\beqar{structure}
\left(\!\not\!{P}C \right)_{\al \be} \left(\ga_5 N\right)_\ga &=&
\left(\!\not\!{p}C \right)_{\al \be} \left(\ga_5 N^+\right)_\ga +
\left(\!\not\!{p}C \right)_{\al \be} \left(\ga_5 N^-\right)_\ga 
\nn \\
&& {}+\frac{M^2}{2 p\cdot z} 
\left(\!\not\!{z}C \right)_{\al \be} \left(\ga_5 N^+\right)_\ga +
\frac{M^2}{2 p\cdot z} 
\left(\!\not\!{z}C \right)_{\al \be} \left(\ga_5 N^-\right)_\ga 
\, .
\eeqar
The first structure on the r.h.s. is of order $(p^+)^{3/2}$, 
the second of order $(p^+)^{1/2}$, the third of 
order $(p^+)^{-1/2}$ and the fourth $(p^+)^{-3/2}$,
respectively. These contributions are interpreted as twist-3, twist-4,
twist-5 and twist-6, respectively, 
 and it follows that the invariant function $\V_1$
contributes to all twists starting from the leading one. Such an effect
is familiar from deep inelastic scattering, where the 
twist-3 structure function $g_2(x,Q^2)$ receives the so-called 
Wandzura-Wilczek contribution related to the leading-twist 
structure function $g_1(x,Q^2)$.

The twist classification based on  counting of powers of $1/p^+$ 
is mathematically similar to the light-cone quantisation approach of 
\cite{Kog70}.
In this language, one decomposes the quark fields contained in the 
matrix element \Gl{zerl} in `plus' and `minus' components 
$q = q^+ + q^-$ in the same manner as done above with the nucleon 
spinor \Gl{spinor}. 
The leading twist amplitude is identified as the one containing
three `plus' quark fields while each `minus' component 
introduces one additional unit of twist. 
Up to possible complications due to isospin, 
one expects, therefore,  to find  eight independent
three quark nucleon distribution amplitudes: One corresponding to
the twist-3 operator $(u^+ u^+ d^+)$,  three related to the possible 
twist-4 operators $(u^+ u^+ d^-), (u^+ u^- d^+), (u^- u^+ d^+)$, 
three more amplitudes of twist-5 of the type 
$(u^- u^- d^+), (u^- u^+ d^-), (u^+ u^- d^-)$
and one amplitude of twist-6 having the structure $(u^- u^- d^-)$.

Either way, distribution amplitudes of definite twist 
correspond to the decomposition of \Gl{zerl} in different light-cone 
components. After a simple algebra, we arrive at the following 
definition of light-cone nucleon distribution amplitudes:  
\beqar{defdisamp}
\lefteqn{ 4 \bra{0} \ep^{ijk} u_\al^i(a_1 z) u_\be^j(a_2 z) d_\ga^k(a_3 z) 
\ket{P} =}
\nn \\
&=&  
S_1 M C_{\al \be} \left(\ga_5 N^+\right)_\ga + 
S_2 M C_{\al \be} \left(\ga_5 N^-\right)_\ga + 
P_1 M \left(\ga_5 C\right)_{\al \be} N^+_\ga + 
P_2 M \left(\ga_5 C \right)_{\al \be} N^-_\ga  
\nn \\
&& + 
V_1  \left(\!\not\!{p}C \right)_{\al \be} \left(\ga_5 N^+\right)_\ga + 
V_2  \left(\!\not\!{p}C \right)_{\al \be} \left(\ga_5 N^-\right)_\ga + 
\frac{V_3}{2} M  \left(\ga_\perp C \right)_{\al \be}\left( \ga^{\perp} 
\ga_5 N^+\right)_\ga 
\nn \\ 
&& +
\frac{V_4}{2} M  \left(\ga_\perp C \right)_{\al \be}\left( \ga^{\perp} 
\ga_5 N^-\right)_\ga + 
V_5  \frac{M^2}{2 p z} 
\left(\!\not\!{z}C \right)_{\al \be} \left(\ga_5 N^+\right)_\ga + 
\frac{M^2}{2 pz} V_6  \left(\!\not\!{z}C \right)_{\al \be} \left(\ga_5 N^-\right)_\ga 
\nn \\
&& + 
A_1  \left(\!\not\!{p}\ga_5 C \right)_{\al \be} N^+_\ga + 
A_2  \left(\!\not\!{p}\ga_5 C \right)_{\al \be} N^-_\ga + 
\frac{A_3}{2} M  \left(\ga_\perp \ga_5 C \right)_{\al \be}\left( \ga^{\perp} 
N^+\right)_\ga 
\nn \\ 
&& +
\frac{A_4}{2} M  \left(\ga_\perp \ga_5 C \right)_{\al \be}\left( \ga^{\perp} 
N^-\right)_\ga + 
A_5  \frac{M^2}{2 p z} 
\left(\!\not\!{z}\ga_5 C \right)_{\al \be} N^+_\ga + 
\frac{M^2}{2 pz}  A_6  \left(\!\not\!{z}\ga_5 C \right)_{\al \be} N^-_\ga 
\nn \\
&& +
T_1 \left(i \si_{\perp p} C\right)_{\al \be} 
\left(\ga^\perp\ga_5 N^+\right)_\ga + 
T_2 \left(i \si_{\perp\, p} C\right)_{\al \be} 
\left(\ga^\perp\ga_5 N^-\right)_\ga + 
T_3 \frac{M}{p z}
\left(i \si_{p\, z} C\right)_{\al \be} 
\left(\ga_5 N^+\right)_\ga  
\nn \\
&& 
+ T_4 \frac{M}{p z}\left(i \si_{z\, p} C\right)_{\al \be} 
\left(\ga_5 N^-\right)_\ga + 
T_5 \frac{M^2}{2 p z}  \left(i \si_{\perp\, z} C\right)_{\al \be} 
\left(\ga^\perp\ga_5 N^+\right)_\ga + 
\frac{M^2}{2 pz}  T_6 \left(i \si_{\perp\, z} C\right)_{\al \be} 
\left(\ga^\perp\ga_5 N^-\right)_\ga  
\nn \\ 
&& + M \frac{T_7}{2} \left(\si_{\perp\, \perp'} C\right)_{\al \be} 
\left(\si^{\perp\, \perp'} \ga_5 N^+\right)_\ga +
M \frac{T_8}{2} \left(\si_{\perp\, \perp'} C\right)_{\al \be} 
\left(\si^{\perp\, \perp'} \ga_5 N^-\right)_\ga\,, 
\eeqar
where an obvious notation $\sigma_{pz} = \sigma^{\mu\nu} p_\mu z_\nu $ 
etc. is used as a shorthand and $\perp$ stands for the projection 
transverse to $z,p$, e.g. 
$\gamma_\perp \gamma^\perp = \gamma^\mu g_{\mu\nu}^\perp \gamma^\nu$ with 
$g_{\mu\nu}^\perp = g_{\mu\nu}-(p_\mu z_\nu+z_\mu p_\nu)/pz$. 

By power counting we identify  three twist-3 distribution amplitudes 
$V_1,A_1,T_1$, nine twist-4 and twist-5, respectively, 
and three  twist-6 distributions, see Table~\ref{tabelle1}.
\begin{table}
\renewcommand{\arraystretch}{1.3}
\begin{center}
\begin{tabular}{l|l|l|l|l}
         &  twist-3  &  twist-4  & twist-5       & twist-6  \\ \hline
vector   &  $V_1$    & $V_2\;,\;V_3$ & $V_4\;,\;V_5 $& $V_6$ \\ \hline
pseudo-vector &  $A_1$    & $A_2\;,\;A_3$ & $A_4\;,\;A_5 $& $A_6$ \\ \hline
tensor  & $T_1$ & $T_2\;,\;T_3\;,T_7$ 
                & $T_4\;,\;T_5\;,T_8$ & $T_6$ \\ \hline
scalar   &      & $S_1$ & $S_2$&   \\ \hline
pseudo-scalar   &      & $P_1$ & $P_2$&   \\ \hline

\end{tabular}
\end{center} 
\caption[]{\sf Twist classification of the distribution amplitudes
in \Gl{defdisamp}}
\label{tabelle1} 
\renewcommand{\arraystretch}{1.0}
\end{table}
Each distribution amplitude $F = V_i,A_i,T_i,S_i,P_i$ can be 
represented as 
\beqar{fourier}
F(a_i p\cdot z) = \int \! {\cal D} x\, e^{-ipz 
\sum_i x_i a_i} F(x_i)\,,
\eeqar
where the functions $F(x_i)$ depend on the dimensionless
variables $x_i,\, 0 < x_i < 1, \sum_i x_i = 1$ which 
correspond to the longitudinal momentum fractions 
carried by the quarks inside the nucleon.  
The integration measure is defined as 
\beqar{integration}
\int\! {\cal D} x  = \int_0^1\! \dd x_1 \dd x_2 \dd x_3\, 
\de (x_1 + x_2 + x_3 - 1)\,.
\eeqar 

Comparision of the expansions in \Gl{zerl} and \Gl{defdisamp} then leads to
 the expressions for the invariant functions 
$\V_i, \A_i, \T_i, {\cal S}_i,{\cal P}_i$ 
in terms of the distribution amplitudes.
For scalar and pseudo-scalar distributions we have:
\beqar{opesp}
\renewcommand{\arraystretch}{1.7}
\hspace*{-4.7cm}
\begin{array}{lll}
 {\cal S}_1 = S_1\,, &\qquad\qquad\qquad& 2\pz\, {\cal S}_2 = S_1-S_2\,, \\
 {\cal P}_1 = P_1\,, &\qquad\qquad\qquad& 2\pz\, {\cal P}_2 = P_2-P_1\,, 
\end{array}
\renewcommand{\arraystretch}{1.0}
\eeqar
for vector distributions:
\beqar{opev}
\renewcommand{\arraystretch}{1.7}
\begin{array}{lll}
 \V_1 = V_1\,, &~~& 2 p\cdot z \V_2 = V_1 - V_2 - V_3\,, \\
 2 \V_3 = V_3\,, &~~& 4 p\cdot z \V_4 = - 2 V_1 + V_3 + V_4  + 2 V_5\,, \\
4 p\cdot z \V_5 = V_4 - V_3\,, &~~&
(2 p\cdot z )^2  \V_6 = - V_1 + V_2 +  V_3 +  V_4 + V_5 - V_6\,,
\end{array}
\renewcommand{\arraystretch}{1.0}
\eeqar
for axial vector distributions:
\beqar{opea}
\renewcommand{\arraystretch}{1.7}
\begin{array}{lll}
 \A_1 = A_1\,, &~~& 2 p\cdot z \A_2 = - A_1 + A_2 -  A_3\,, \\
 2 \A_3 = A_3\,, &~~& 4 p\cdot z \A_4 = - 2 A_1 - A_3 - A_4  + 2 A_5\,, \\
 4 p\cdot z \A_5 = A_3 - A_4\,, &~~&
(2 p\cdot z )^2  \A_6 =  A_1 - A_2 +  A_3 +  A_4 - A_5 + A_6\,,
\end{array}
\renewcommand{\arraystretch}{1.0}
\eeqar
and, finally, for tensor distributions:
\beqar{opet}
\renewcommand{\arraystretch}{1.7}
\begin{array}{lll}
 \T_1 = T_1\,, &~~& 2 \pz \T_2 = T_1 + T_2 - 2 T_3\,, \\
2 \T_3 = T_7\,, &~~& 2 \pz \T_4 = T_1 - T_2 - 2  T_7\,, \\
2 \pz \T_5 = - T_1 + T_5 + 2  T_8\,, &~~&
    (2 \pz)^2 \T_6 = 2 T_2 - 2 T_3 - 2 T_4 + 2 T_5 + 2 T_7 + 2 T_8\,,\\
4 \pz \T_7 = T_7 - T_8\,, &~~&
(2 \pz)^2 \T_8 = -T_1 + T_2 + T_5 - T_6 + 2 T_7 + 2 T_8 \,.
\end{array}
\renewcommand{\arraystretch}{1.0}
\eeqar

\subsection{Symmetry properties}

Not all of the 24 distribution amplitudes in \Gl{defdisamp} are independent.
First of all, each distribution amplitude itself has definite
symmetry properties. 
The identity of the two $u$-quarks in the 
proton together with symmetry properties of quark operators and 
$\ga$-matrices implies that the vector and tensor distribution 
amplitudes are symmetric, whereas scalar, pseudoscalar and axial-vector 
distributions are antisymmetric under the interchange of the 
first two arguments:
\beqar{symm}
V_i(1,2,3) =  \phantom{-}V_i(2,1,3) \,, 
&\qquad&  
T_i(1,2,3) =  \phantom{-}T_i(2,1,3) \,,
\\
S_i(1,2,3) = - S_i(2,1,3) \,, &\qquad& P_i(1,2,3) =  - P_i(2,1,3)\,, 
\hspace{1cm}
A_i(1,2,3) = - A_i(2,1,3) \,.
\nn
\eeqar
Similar relations hold for the ``calligraphic'' structures in \Gl{zerl}.   

In addition, the full matrix element in \Gl{defdisamp} has to fulfill 
the symmetry relation  
\begin{equation}
\bra{0} \ep^{ijk} u_\al^i(1) u_\be^j(2) d_\ga^k(3) \ket{P} +
\bra{0} \ep^{ijk} u_\al^i(1) u_\ga^j(3) d_\be^k(2 ) \ket{P} +
\bra{0} \ep^{ijk} u_\ga^i(3) u_\be^j(2) d_\al^k(1) \ket{P} = 0 \,.
\label{isospin}
\end{equation}
that follows from the condition that the nucleon state has isospin 1/2:
\beqar{operator}
&&\left(T^2 - \frac{1}{2}(\frac{1}{2} +1)\right)
\bra{0} \ep^{ijk} u_\al^i(1) u_\be^j(2) d_\ga^k(3) 
\ket{P} = 0\,,
\eeqar 
where 
\beqar{op2}
T^2 = \frac{1}{2} \left(T_+T_- + T_- T_+ \right) + T_3^2 \,
\eeqar
and $T_\pm$ are the usual isospin step-up and step-down operators. 
Applying the set of Fierz transformations detailed in Appendix~A,
we end up with the following six equations:
\beqar{isospin2}
2 T_1(1,2,3) &=& [V_1-A_1](1,3,2) + [V_1-A_1](2,3,1)\,, \nn  \\
\left[T_3+   T_7 + S_1-P_1\right] (1,2,3) 
&=&   [V_3 - A_3](3,1,2) + [V_2 - A_2](2,3,1)\,, \nn \\
2 T_2(1,2,3) &=&   [T_3 -  T_7 + S_1 + P_1](3,1,2)  + 
[T_3 -  T_7 + S_1 + P_1](3,2,1)\,, \nn \\
\left[ T_4+  T_8 + S_2-P_2\right] (1,2,3) &=& 
 [V_4 - A_4](3,1,2) + [V_5 - A_5](2,3,1)\,, \nn \\
2 T_5(1,2,3) &=&   [T_4 -  T_8 + S_2 + P_2](3,1,2)  +
[T_4 -  T_8 + S_2 + P_2](3,2,1)\,, \nn \\
2 T_6(1,2,3) &=& [V_6-A_6](1,3,2) + [V_6-A_6](2,3,1)\,. 
\eeqar
The first relation in \Gl{isospin2} is familiar from \cite{Che84}
and expresses the tensor nucleon distribution amplitude of the leading   
twist in terms of the vector and axial vector distributions. Since the 
latter have different symmetry, they can be combined together to define 
the single independent leading twist-3 nucleon distribution amplitude
\beqar{twist-3}
\Phi_3(x_1,x_2,x_3) = [V_1 - A_1](x_1,x_2,x_3) 
\eeqar
which is well known and received a lot of attention in the past.
The rest of the relations in \Gl{isospin2} are new and constrain the 
number of independent distribution amplitudes of higher twist.

In particular, nine distributions amplitudes of twist-4 (cf. Table~1)
can be reduced to three independent distributions that we choose 
to be 
\beqar{twist-4}
\Phi_4(x_1,x_2,x_3) &=& [V_2 - A_2](x_1,x_2,x_3) \,, \quad \nn \\
\Psi_4(x_1,x_2,x_3) &=& [V_3 - A_3](x_1,x_2,x_3) \,, \nn \\ 
\Xi_4(x_1,x_2,x_3) &=& [T_3 -  T_7 +  S_1 + P_1](x_1,x_2,x_3)
\,.
\eeqar
It is easy to check that all nine distributions 
$V_2, A_2, T_2, V_3, A_3, T_3, T_7, S_1, P_1$ can be restored from 
the three above, taking into account the isospin relations in the second and
the third lines in \Gl{isospin2}.

Similarly, we introduce three independent twist-5 distribution amplitudes
\beqar{twist-5}
\Phi_5(x_1,x_2,x_3) &=& 
[V_5 - A_5](x_1,x_2,x_3) \, \,  , \quad \nn \\
\Psi_5(x_1,x_2,x_3) &=& [V_4 - A_4](x_1,x_2,x_3)  \, ,\nn \\ 
\Xi_5(x_1,x_2,x_3) &=& [T_4 -  T_8 +  S_2 + P_2](x_1,x_2,x_3)
\,. 
\eeqar
Finally, one distribution amplitude exists of twist-6: 
\beqar{twist-6}
\Phi_6(x_1,x_2,x_3) = [V_6 - A_6](x_1,x_2,x_3)\,. 
\eeqar

\subsection{Representation in terms of chiral fields}

The physics interpretation of the distribution amplitudes $\Phi(x_i)$
is most transparent in terms of quark fields of definite chirality 
\beqar{chiral}
q^{\up(\down)} = \frac{1}{2} (1 \pm \ga_5) q \,.
\eeqar
Projection on the state where the spins of the two up-quarks are
antiparallel $u^\up u^\down$ singles out vector and axial vector amplitudes, 
while  parallel spins  $u^\up u^\up$ or $u^\down u^\down$ correspond 
to scalar, pseudoscalar and tensor structures.
We are lead to the classification of the distribution amplitudes in terms
of the quarks light-cone components and spin projections, summarized in 
Table~\ref{tabelle2}.
\begin{table}
\renewcommand{\arraystretch}{1.3}
\begin{center}
\begin{tabular}{l|l|l|l}
&  
Lorentz-structure 
&  
Light-cone projection   
& 
nomenclature
\\ \hline
twist-3   
&  $ \left(C\!\not\!{z}\right) \otimes \!\not\!{z} $    
& $u^+_\up u^+_\down d^+_\up$
& $\Phi_3(x_i) = \left[V_1-A_1\right](x_i)$ \\ \hline
twist-4
&  $ \left(C\!\not\!{z}\right) \otimes \!\not\!{p} $    
& $u^+_\up u^+_\down d^-_\up$
& $\Phi_4(x_i) = \left[V_2-A_2\right](x_i)$ 
\\ \hline
& $ \left(C\!\!\not\!{z}\ga_\perp\!\!\not\!{p}\,\right) 
      \otimes \ga^\perp\!\!\not\!{z} $    
& $ u^+_\up u^-_\down d^+_\down$
& $\Psi_4(x_i) = \left[V_3-A_3\right](x_i)$ \\ \hline
& $ \left(C \!\not\!{p}\!\not\!{z}\right)  \otimes \!\not\!{z}$    
& $u^-_\up u^+_\up  d^+_\up$
& $\Xi_4(x_i)
= \left[T_3-  T_7 + S_1 + P_1\right](x_i)$ \\ \hline
twist-5
&  $ \left(C\!\not\!{p}\right) \otimes \!\not\!{z} $    
& $u^-_\up u^-_\down d^+_\up$
& $\Phi_5(x_i) = \left[V_5-A_5\right](x_i)$ 
\\ \hline
& $ \left(C\!\!\not\!{p}\ga_\perp\!\!\not\!{z}\,\right) 
\otimes \ga^\perp\!\!\not\!{p} $    
& $ u^-_\up u^+_\down d^-_\down
$
& $\Psi_5(x_i) = \left[V_4-A_4\right](x_i)$ \\ \hline
& $ \left(C \!\not\!{z}\!\not\!{p}\right)  \otimes \!\not\!{p}$    
& $u^+_\up u^-_\up  d^-_\up$
& $\Xi_5(x_i)
= \left[T_4-  T_8 + S_2 + P_2\right](x_i)$ \\ \hline
twist-6   
&  $ \left(C\!\not\!{p}\right) \otimes \!\not\!{p} $    
& $u^-_\up u^-_\down d^-_\up$
& $\Phi_6(x_i) = \left[V_6-A_6\right](x_i)$ \\ \hline
\end{tabular}
\end{center} 
\caption[]{\sf 
 Eight independent nucleon distribution amplitudes 
that enter the expansion in \Gl{defdisamp}.}
\label{tabelle2} 
\renewcommand{\arraystretch}{1.0}
\end{table}
The leading twist-3 distribution amplitude can be defined as 
(cf. \cite{Bra99}):
\beqar{vector-twist-3}
\hspace*{-1cm}
\bra{0} \ep^{ijk}\! 
\left(u^{\up}_i(a_1 z) C \!\!\not\!{z} u^{\down}_j(a_2 z)\right)  
\!\not\!{z} d^{\up}_k(a_3 z) \ket{P} 
&=& - \frac12 pz \!\not\!{z} N^\up\! \!\int\!\! \D x 
\,e^{-i pz \sum x_i a_i}\, 
\Phi_3(x_i)\,.
\eeqar
Twist-4 distributions allow for the following representation:
\beqar{vector-twist-4}
\hspace*{-0.3cm}\bra{0} \ep^{ijk}\! 
\left(u^{\up}_i(a_1 z) C \!\!\not\!{z} u^{\down}_j(a_2 z)\right) 
\!\not\!{p} d^{\up}_k(a_3 z) \ket{P}
&=& - \frac12 pz \!\not\!{p} N^\up\!\! \int\! \D x 
\,e^{-i pz \sum x_i a_i}\, 
\Phi_4(x_i)\,, 
\nn \\
\hspace*{-0.3cm}\bra{0} \ep^{ijk}\! 
\left(u^{\up}_i(a_1 z) C\!\not\!{z}\ga_\perp\!\!\not\!{p} 
 u^{\down}_j(a_2 z)\right) 
\ga^\perp\! \!\!\not\!{z} d^{\down}_k(a_3 z) \ket{P}
&=&   - pz M \!\!\not\!{z} N^\up\! \int\!\! \D x 
\,e^{-i pz \sum x_i a_i}\, 
\Psi_4(x_i)\,,
 \nn\\  
\hspace*{-0.3cm}\bra{0} \ep^{ijk}\! 
\left(u^{\up}_i(a_1 z) C \!\not\!{p}\!\!\not\!{z} u^{\up}_j(a_2 z)\right)  
\!\not\!{z} d^{\up}_k(a_3 z) \ket{P}
&=&  \frac12 pz M \!\not\!{z} N^\up\! \int\!\! \D x 
\,e^{-i pz \sum x_i a_i}\, 
\Xi_4(x_i) 
\eeqar
and, similar, for twist-5:
\beqar{vector-twist-5}
\hspace*{-0.3cm}\bra{0} \ep^{ijk}\! 
\left(u^{\up}_i(a_1 z) C \!\not\!{p} u^{\down}_j(a_2 z)\right) 
\!\not\!{z} d^{\up}_k(a_3 z) \ket{P}
&=& - \frac14 M^2 \!\not\!{z} N^\up\! \int\!\! \D x 
\,e^{-i pz \sum x_i a_i}\, 
\Phi_5(x_i)\,, 
\nn \\ 
\hspace*{-0.3cm}\bra{0} \ep^{ijk}\! 
\left(u^{\up}_i(a_1 z) C \!\!\not\!{p}\ga_\perp\!\!\not\!{z}
 u^{\down}_j(a_2 z)\right) 
\ga^\perp \!\!\not\!{p}\, 
d^{\down}_k(a_3 z) \ket{P}
&=&   -  pz M  \!\not\!{p}\, N^\up \int \D x 
\,e^{-i pz \sum x_i a_i}\, 
\Psi_5(x_i)\,,
\nn \\  
\hspace*{-0.3cm}\bra{0} \ep^{ijk}\! 
\left(u^{\up}_i(a_1 z) C\!\!\not\!{z}\!\!\not\!{p}\, u^{\up}_j(a_2 z)\right)  
\!\not\!{p} d^{\up}_k(a_3 z) \ket{P}
&=&  \frac12 pz M \!\not\!{p}\, N^\up\! \int\!\! \D x 
\,e^{-i pz \sum x_i a_i}\, 
\Xi_5(x_i)\,. 
\eeqar
Finally, the twist-6 distribution amplitude is written as
\beqar{vector-twist-6}
\hspace*{-0.3cm}\bra{0} \ep^{ijk}\! 
\left(u^{\up}_i(a_1 z) C \!\!\not\!{p} u^{\down}_j(a_2 z)\right)  
\!\not\!{p} d^{\up}_k(a_3 z) \ket{P}
&=& - \frac14 M^2 \!\not\!{p} N^\up\! \int\!\! \D x 
\,e^{-i pz \sum x_i a_i}\,
\Phi_6(x_i)\,.
\eeqar

In the rest of the paper we study this set of  distribution amplitudes 
in some more detail. Final expressions for all the 24 
distribution amplitudes in \Gl{defdisamp} are collected in Appendix~C.
 
\section{Conformal expansion}
\setcounter{equation}{0}

The conformal expansion of light-cone distribution amplitudes
is the field-theoretic analogue to the partial wave expansion in 
quantum mechanics. The idea, in both cases, is to use the symmetry 
of the problem to introduce a set of separated coordinates. 
In quantum mechanics, spherical symmetry of the 
potential allows to separate dependence on radial coordinates from 
angular ones. All angular dependence is included in 
spherical harmonics which form an irreducible representation of the 
symmetry group O(3), while the dependence on radial coordinates 
is governed by a one-dimensional Schr\"odinger equation.

In the same spirit conformal symmetry \cite{MS69} of the QCD Lagrangian 
can be used to  study the distribution amplitudes as it allows to
separate longitudinal degrees of freedom  from transverse 
ones \cite{B+,Makeenko,O82,BB88,Bra90,Mueller,Bra99}.
The dependence on longitudinal momentum fractions is   
taken into account by a set of orthogonal polynomials
that form an irreducible representation of the collinear 
subgroup $SL(2,R)$ of the conformal group 
which describes M\"obius transformations on the light-cone.
Transverse coordinates are replaced by the renormalization scale, the 
dependence on which is governed by the renormalization group.
Since the renormalization group equations to leading logarithmic accuracy 
are driven by tree-level counterterms, they have the conformal symmetry.
As a consequence, components in the distribution amplitudes with 
different conformal spin, dubbed conformal partial waves, do not 
mix under renormalization to this accuracy.

Conformal spin of the quark is defined as 
\beqar{confspin}
j = \frac{1}{2} (l+s)
\eeqar 
where $l=3/2$ is the canonical dimension of the quark field and $s=\pm 1/2$
is the quark spin projection on the light-cone. The spin projection 
operators are in fact the same as used to separate the ``plus'' and ``minus''
components of a spinor in \Gl{project}. The ``plus'' component of the 
quark field $q^+$ corresponds to $s=1/2$ and, therefore,  $j=1$, while 
the ``minus'' component $q^-$ has $s=-1/2$ and $j=1/2$.
For multiquark states, we are left with the classical problem of summation
of spins, with the difference that the group is in our case non-compact.
The distribution amplitude corresponding to the lowest conformal spin 
$j_{\rm min} = j_1 + j_2 + j_3$ of the three-quark system is equal to 
\cite{BB88,Bra90}
\beqar{asymptotic}
\Phi_{\rm as}(x_1,x_2,x_3)  = 
\frac{\Gamma[2 j_1 + 2 j_2 + 2 j_3]}{\Gamma[2 j_1]\Gamma[2 j_2]\Gamma[2 j_3]}
x_1^{2j_1 -1} x_2^{2j_2 -1} x_3^{2j_3 -1}\,.
\eeqar
Contributions with higher conformal spin $j=j_{\rm min}+n, n=1,2,\ldots$
are given by $\Phi_{\rm as}$ multiplied by polynomials which are 
orthogonal over the weight function \Gl{asymptotic}. 
A suitable orthonormal basis of such 
``conformal polynomials'' has been constructed in \cite{Bra99}.
   
In this section we consider the conformal expansion of nucleon 
distribution amplitudes taking into account contributions of leading and 
next-to-leading conformal spin (i.e. ``S'' and ``P''- waves).
The expansions can easily be extended to arbitrary spin. One reason why 
we do not present complete expansions in this paper is that we 
do not have the tools to estimate the corresponding additional parameters.
The second reason is that for yet higher spins one has to take into 
account contributions of four-particle distributions involving an 
extra gluon that we do not consider here.

\subsection{Leading twist-3 distribution amplitude} 

At leading twist, there is only one independent distribution amplitude 
corresponding to the light-cone projection
$ \left(C\!\not\!{z}\right) \otimes \!\not\!{z} $ and therefore 
involving three ``plus'' quark fields , see Table~2.
The conformal expansion then  reads
\beqar{exp-twist-3-vector} 
\Phi_3(x_i,\mu) &=& 120 x_1 x_2 x_3 \left[\phi_3^0(\mu) + 
  \phi_3^-(\mu) (x_1 - x_2) 
+ \phi_3^+(\mu) (1- 3 x_3)+\ldots\right]\,. 
\eeqar
Here $\mu$ is the renormalization scale. For the discussion of the shape 
of the distribution it is often convenient to rewrite 
\Gl{exp-twist-3-vector} to factor out the overall normalization: 
\beqar{exp-twist-3-vector-b} 
\Phi_3(x_i,\mu) &=& 120 x_1 x_2 x_3 \phi_3^0(\mu) 
\left[1 + \widetilde\phi_3^-(\mu) (x_1 - x_2) 
+ \widetilde\phi_3^+(\mu) (1- 3 x_3)+\ldots\right]\,. 
\eeqar
The relation between $\phi_3^\pm$ and $\widetilde\phi_3^\pm$ is obvious.

The statement of conformal 
symmetry is that the coefficients $\phi^\pm(\mu)$ 
do not mix  with $\phi^0(\mu)$ under renormalization since
they have different conformal spin:  $j=4$ and $j=3$, respectively. 
By an explicit calculation one obtains \cite{LB79,P79,T82,Nyeo}
\beqar{t3-renorm}
     \phi_3^{0}(\mu_2) &=& L^{2/3b} \phi_3^{0}(\mu_1)\,,\qquad
     L \equiv \frac{\alpha_s(\mu_2)}{\alpha_s(\mu_1)}\,,
\eeqar 
where $b= 11 - 2/3 n_f$, and  
\beqar{t3-renorm2}
  \widetilde\phi_3^{+}(\mu_2) &=& 
\frac14 \left( 3 L^{20/9b}+ L^{8/3b}\right)\widetilde \phi_3^{+}(\mu_1)
+\frac14 \left( L^{20/9b}- L^{8/3b}\right)\widetilde\phi_3^{-}(\mu_1)\,,
\nn\\
 \widetilde\phi_3^{-}(\mu_2) &=& 
\frac34 \left( L^{20/9b}- L^{8/3b}\right)\widetilde\phi_3^{+}(\mu_1)
+\frac14 \left( L^{20/9b}+  3 L^{8/3b}\right)\widetilde\phi_3^{-}(\mu_1)\,.
\eeqar
Numerical estimates for the coefficients are available from QCD sum 
rules~\cite{Che84}:\footnote{In notations of \cite{Che84} 
$\phi_3^{0} \equiv f_N$. The given numbers correspond to the last 
reference in \cite{Che84}.}
\beqar{twist3-numerical}
  \phi_3^{0}(\mu = 1~{\rm GeV}) &=& (5.3\pm 0.5)\cdot 10^{-3}~{\rm GeV}^2\,,  
\nn\\
    \widetilde\phi_3^{+}(\mu = 1~{\rm GeV}) &=& 1.1 \pm 0.3\,,
\nn\\
    \widetilde\phi_3^{-}(\mu = 1~{\rm GeV}) &=& 4.0 \pm 1.5\,.
\eeqar
More sophisticated models suggested in \cite{Che84} involve in addition
contributions of second order polynomials related to the operators 
with conformal spin-5. The estimates of the corresponding coefficients are,
however, less reliable and have large errors.

\subsection{Higher-twist distribution amplitudes}

The conformal expansion of the higher-twist distribution amplitudes 
defined in Sect.~2 is equally straightforward. With the help 
of the general expression in \Gl{asymptotic} one obtains for twist-4:
\beqar{twist4-expand} 
\Phi_4(x_i) &=& 24 x_1 x_2 
\left[\phi_4^0 + \phi_4^- (x_1 - x_2) + 
\phi_4^+ (1- 5 x_3)\right] \,,
\nn\\
\Psi_4(x_i) &=& 24 x_1 x_3 
\left[\psi_4^0 + \psi_4^- (x_1 - x_3) + 
\psi_4^+ (1- 5 x_2)\right]\,,
\nn\\
\Xi_4(x_i) &=& 24 x_2 x_3 
\left[\xi_4^0  + \xi_4^-  (x_2 - x_3) + 
\xi_4^+(1- 5 x_1)\right] \,,
\eeqar
for twist-5:
\beqar{twist5-expand}
\Phi_5(x_i) &=& 6 x_3 
\left[\phi_5^0  + \phi_5^- (x_1 - x_2) + \phi_5^+ (1- 2 x_3)\right]\,, 
\nn\\
\Psi_5(x_i) &=& 6 x_2
\left[\psi_5^0 + \psi_5^- (x_1 - x_3) + \psi_5^+(1- 2 x_2)\right]\,,
\nn\\
\Xi_5(x_i) &=& 6 x_1 
\left[\xi_5^0  + \xi_5^-  (x_2 - x_3) + \xi_5^+ (1- 2 x_1)\right] \,,
\eeqar 
and for twist-6:
\beqar{twist6-expand} 
\Phi_6(x_i) &=& 2 \left[\phi_6^0 + \phi_6^- (x_1 - x_2) 
+ \phi_6^+ (1- 3 x_3)\right]\,. 
\eeqar
At this point the expansion introduces 21 new parameters. 
Our next task is to find out how many of the parameters 
actually are independent and which are connected by equations
of motion.

The normalisation of all distribution amplitudes and, therefore, 
the asymptotic wave functions are determined by matrix elements of 
a local three-quark operator without derivatives. 
The Lorenz decompositon of a local three quark matrix element 
is much simpler compared to the general parametrisation in \Gl{zerl}
and involves only four structures:
\beqar{zerl0}
&& 
4 \bra{0} \ep^{ijk} u_\al^i(0) u_\be^j(0) d_\ga^k(0) \ket{N,P}
=   
{\cal V}^0_1  \left(\!\not\!{P}C \right)_{\al \be} \left(\ga_5 N\right)_\ga + 
\V^0_3 M  \left(\ga_\mu C \right)_{\al \be}\left( \ga^{\mu} \ga_5 N\right)_\ga 
\nn \\ &&
\T^0_1 \left(P^\nu i \si_{\mu\nu} C\right)_{\al \be} 
\left(\ga^\mu\ga_5 N\right)_\ga 
+
\T^0_3 M \left(\si_{\mu\nu} C\right)_{\al \be} 
\left(\si^{\mu\nu}\ga_5 N\right)_\ga  \; .
\eeqar
From isospin constraints it follows that in addition $\V^0_1 = \T^0_1$. 
Thus there exist only three independent constants.

Remarkably, these three parameters are well known and can be obtained 
from existing estimates of the following three matrix elements:
\beqar{local-operators} 
&& \bra{0} \ep^{ijk} \left[u^i(0) C \!\not\!{z} u^j(0)\right] \,
\ga_5 \!\not\!{z} d^k(0) \ket{P}
=   f_{\rm N}
pz \!\not\!{z} N(P)\,,  \nn \\
&&\bra{0} \ep^{ijk} \left[u^i(0) C\ga_\mu u^j(0)\right]\, 
\ga_5 \ga^\mu d^k(0) \ket{P}
= \la_1 M N(P) \,,  
\nn \\
&&\bra{0} \ep^{ijk} \left[u^i(0) C\si_{\mu\nu} u^j(0)\right] 
\, \ga_5 \si^{\mu\nu} d^k(0) \ket{P}
= \la_2 M N(P) \,.  
\eeqar
The parameter $f_{\rm N} = \V_1^0$ enters already at the level of twist-3 and 
determines the normalisation of the leading twist distribution amplitude 
\Gl{twist3-numerical}. 
The other two parameters $\la_1 = (\V_1^0 - 4 \V_3^0)$ and 
$\la_2 = 6 (\V_1^0 - 4 \T_3^0)$  correspond to the nucleon coupling to 
the two possible independent nucleon interpolating fields that are
widely used in calculations of dynamical
characteristics of the nucleon in the QCD sum rule approach.
The operator corresponding to $\la_1$ was introduced in \cite{Iof81}
while the one corresponding to $\la_2$ was advertised in \cite{Chu82,Kol84}.
The QCD sum rules summarized in App.~\ref{app:b} yield the following
estimates:
\beqar{twist4-numerical}
  \lambda_1(\mu = 1~{\rm GeV}) &=& - (2.7\pm 0.9)\cdot 10^{-2}~{\rm GeV}^2\,,  
\nn\\
  \lambda_2(\mu = 1~{\rm GeV}) &=& \phantom{-}
                         (5.1 \pm 1.9)\cdot 10^{-2}~{\rm GeV}^2\,,  
\eeqar
Anomalous dimensions are the same for both currents \cite{Piv91}
\beqar{t4-renorm}
     \la_1(\mu_2) = L^{2/b} \la_1(\mu_1)\,,\qquad
     \la_2(\mu_2) = L^{2/b} \la_2(\mu_2)\,.\qquad 
\eeqar 
Note that one overall sign in the determination of vacuum-to-nucleon 
matrix elements is arbitrary as it corresponds to arbitrary (unphysical)
overall phase of the nucleon wave function. The relative signs between
the couplings  
$f_N$, $\la_1$, $\la_2$, etc. are, however, well defined and can be 
determined from suitable nondiagonal correlation functions, see
App.~\ref{app:b} for details. We choose $f_{\rm N}$ to be real and positive;
then, it turns out that $\la_1$ is negative and $\la_2$ positive.
The fact that  $\la_1$ and $\la_2$ have opposite signs 
is known from \cite{Kol84}, the negative relative sign 
between $\la_1$ and $f_{\rm N}$ is a new result.

The  coefficients in 
Eqs.~(\ref{twist4-expand}), (\ref{twist5-expand}), (\ref{twist6-expand}) 
corresponding to the operators of leading conformal spin then read
\beqar{local-coefficients}
\phi_3^0 = \phi_6^0 = f_N \,,\hspace{0.3cm} 
&\qquad&  
\phi_4^0 = \phi_5^0 = 
\frac{1}{2} \left(\la_1 + f_N\right) \,, 
\nn \\
\xi_4^0 = \xi_5^0 = 
\frac{1}{6} \la_2\,,
&\qquad&  
\psi_4^0  = \psi_5^0 =          
\frac12\left(f_N - \la_1 \right)  \,.
\nn
\eeqar
Note that the normalization of twist-3 and twist-6 distribution amplitudes
are equal, and similarly for twist-4 and twist-5 distributions. 
The constants $\phi^0_{4,5}$ and $\psi^0_{4,5}$ involve both 
the leading twist matrix element $f_N$ \Gl{local-operators} and the 
higher-twist contribution $\sim \lambda_1$. The former 
is analogous to the Wandzura-Wilczek-type  contribution to the 
higher-twist distribution amplitudes studied in \cite{BallBraun}
for the $\rho$-meson, and the latter defines the ``genuine'' higher-twist
correction. Note that $\lambda_1,\lambda_2 \gg f_N$ 
that is in agreement with our 
expectation (see Introduction) that the matrix elements of higher-twist
three-quark operators are large.   

The remaining contributions of next-to-leading conformal spin
are related to operators with one derivative. In much the same
way as before and as elaborated in App.~\ref{app:b}
 the 14 unknown parameters are reduced by isospin symmetry and 
equations of motion to 
five dimensionless parameters $V_1^d,A_1^u, f_1^d,f_2^d,f_1^u$ 
which we define in the following way:
\beqar{local-operators-2} 
\bra{0} \left(u(0) C \!\not\!{z} u(0)\right) \,
\ga_5 \!\not\!{z} (i z \vecr{D} d)(0) \ket{P}
&=&  \phantom{-}f_{\rm N} V_1^d
(pz)^2 \!\not\!{z} N(P)\,,  \nn \\
\bra{0} \left(u(0) C \!\not\!{z}\ga_5 i z\Dlr u(0)\right) \,
\!\not\!{z} d(0) \ket{P}
&=& - f_{\rm N} A_1^u
(pz)^2 \!\not\!{z} N(P)\,,  \nn \\
\bra{0} \left(u(0) C\ga_\mu u(0)\right)\, 
\ga_5 \ga^\mu \!\not\!{z} (i z \vecr{D} d)(0) \ket{P}
&=& \phantom{-}\la_1 f_1^d (pz) M \!\not\!{z} N(P)\,,  
\nn \\
\bra{0} \left(u^a(0) C\si_{\mu\nu} u(0)\right) 
\, \ga_5 \si^{\mu\nu} \!\not\!{z} 
(i z \vecr{D} d)(0) \ket{P}
&=& - \la_2 f_2^d (pz) M \!\not\!{z} N(P)\,,  
\nn \\
\bra{0}\left(u(0) C \ga_\mu \ga_5 i z\Dlr u(0)\right) \,
\ga^\mu \!\not\!{z} d(0) \ket{P}
&=& - \la_1 f_1^u (pz) M \!\not\!{z} N(P)\,. 
\eeqar
Here, we have used the shorthand notation for the left-right derivative
$ i z\cdot\Dlr = i z\cdot (\vecr{D} - \vecl{D})$ and for brevity 
omitted colour indices. 
The first two matrix elements $V_1^d$ and $A_1^u$ are leading
twist-3 and were estimated in \cite{Che84}\footnote{In notation
of \cite{Che84} $V_1^d = V^{(0,0,1)}, A_1^u = 2 A^{(0,1,0)}$}:
\beqar{twist3-numerical-2}
   V_1^d(\mu = 1~{\rm GeV}) &=& 0.23\pm 0.03 \,,  
\nn\\
   A_1^u(\mu = 1~{\rm GeV}) &=& 0.38 \pm 0.15 \,.
\eeqar
We have used these estimates above in \Gl{twist3-numerical}: 
$\widetilde\phi_3^- = \frac{21}{2} A_1^u$ and 
$\widetilde\phi_3^+ = \frac{7}{2} (1 - 3 V_1^d)$.
The other three parameters are genuine higher-twist and are 
estimated in App.~\ref{app:b} using QCD sum rules. We obtain:
\beqar{twist4-numerical-2}
  f_1^d(\mu = 1~{\rm GeV}) &=&  0.6\pm 0.2  \,,
\nn\\
  f_2^d(\mu = 1~{\rm GeV}) &=&  0.15\pm 0.06 \,,  
\nn\\
  f_1^u(\mu = 1~{\rm GeV}) &=&  0.22\pm 0.15 \,.
\eeqar
The remaining coefficients expressed in terms of the above 
parameters read for twist-4:
\beqar{twist4-parameter}
\phi_4^- &=& \frac{5}{4} \left(\la_1(1- 2 f_1^d -4 f_1^u) 
+ f_N( 2 A_1^u - 1)\right) \,,
\nn \\
\phi_4^+ &=& \frac{1}{4} \left( \la_1(3- 10 f_1^d) 
- f_N( 10  V_1^d - 3)\right)\,,
\nn \\
\psi_4^- &=& - \frac{5}{4} \left(\la_1(2- 7 f_1^d + f_1^u) 
+ f_N(A_1^u + 3 V_1^d - 2)\right) \,,
\nn \\
\psi_4^+ &=& - \frac{1}{4} \left(\la_1 (- 2 + 5 f_1^d + 5 f_1^u) 
+ f_N( 2 + 5 A_1^u - 5 V_1^d)\right)\,,
\nn \\
\xi_4^- &=& \frac{5}{16} \la_2(4- 15 f_2^d)\,,
\nn \\
\xi_4^+ &=& \frac{1}{16} \la_2 (4- 15 f_2^d)\,.
\eeqar
for twist-5: 
\beqar{twist5-parameter}
\phi_5^- &=& \frac{5}{3} \left(\la_1(f_1^d - f_1^u) 
+ f_N( 2 A_1^u - 1)\right) \,,
\nn \\
\phi_5^+ &=& - \frac{5}{6} \left(\la_1 (4 f_1^d - 1) 
+ f_N( 3 + 4   V_1^d)\right)\,,
\nn \\
\psi_5^- &=& \frac{5}{3} \left(\la_1 (f_1^d - f_1^u) 
+ f_N( 2 - A_1^u - 3 V_1^d)\right)\,,
\nn \\
\psi_5^+ &=& -\frac{5}{6} \left(\lambda_1 (- 1 + 2 f_1^d + f_1^u) 
+ f_N( 5 + 2 A_1^u -2 V_1^d)\right)\,,
\nn \\
\xi_5^- &=& - \frac{5}{4} \la_2 f_2^d\,,
\nn \\
\xi_5^+ &=& \phantom{-}\frac{5}{36} \lambda_2 (2- 9 f_2^d) \,,
\eeqar
and for twist-6: 
\beqar{twist6-parameter}
\phi_6^- &=& \phantom{-}\frac{1}{2} \left(\la_1 (1- 4 f_1^d - 2 f_1^u) 
+ f_N(1 +  4 A_1^u )\right) \,,
\nn \\
\phi_6^+ &=& - \frac{1}{2}\left(\la_1  (1 - 2 f_1^d) 
+ f_N ( 4 V_1^d - 1)\right)\,.
\eeqar
With these relations, the construction of higher-twist nucleon distribution
amplitudes is complete to our accuracy. Note that numerical values 
of the parameters given in Eqs.~(\ref{twist3-numerical-2}),
(\ref{twist4-numerical}) 
are strongly correlated within the QCD sum rule approach,
and the given errors should not be added together in 
the sums (\ref{twist4-parameter})--(\ref{twist6-parameter}). We expect that
the accuracy of the QCD sum rule calculation of all coefficients
in the distribution amplitudes is of the order of 50\%.
\begin{table}
\renewcommand{\arraystretch}{1.3}
\begin{center}
\begin{tabular}{l||l|l|l||l|l|l||l|l|l|l}
         &  $\phi_i^0$  &  $\phi_i^-$ &  $\phi_i^+$    
         &  $\psi_i^0$  &  $\psi_i^-$ &  $\psi_i^+$    
         &  $\xi_i^0$   &  $\xi_i^-$  &  $\xi_i^+$    
\\ \hline
twist-3: $i = 3$  
         & $\phantom{-}0.53$         &  $\phantom{-}2.11$       &   0.57 
         &              &             &   
         &              &             &  
\\ \hline
twist-4:  $ i = 4$ 
         & $-1.08$        &  $\phantom{-} 3.22 $     &   2.12
         & 1.61         &   $-6.13$     &   0.99
         & 0.85         &   $\phantom{-} 2.79$     &  0.56
\\ \hline
twist-5: $i = 5$  
         & $-1.08$        &  $-2.01$      &   1.42 
         & 1.61         &  $-0.98$      &   -0.99
         & 0.85         &  $-0.95$      &   0.46
\\ \hline
twist-6: $i = 6$  
         & $\phantom{-}0.53$         &  $\phantom{-} 3.09$      &   -0.25 
         &              &             &   
         &              &             &  
\\ \hline
\end{tabular}
\end{center} 
\caption[]{\sf Numerical values for the expansion parameters defined
in \Gl{twist4-expand} to \Gl{twist6-expand}. The values are given
in units of $10^{-2}\,{\rm GeV^2}$.}
\label{tabelle3} 
\renewcommand{\arraystretch}{1.0}
\end{table}
In table \Ta{tabelle3} we have collected the numerical values for 
the expansion parameters.

In \Ab{figure1} we have presented the leading twist distribution
amplitude $\Phi_3(x_i)$, in \Ab{figure2} the three distribution amplitudes
of twist-4 are plotted. Note that the scale on the $z$-axis is not
the same for each plot.

\begin{figure}[h]
\centerline{
   \epsfig{file=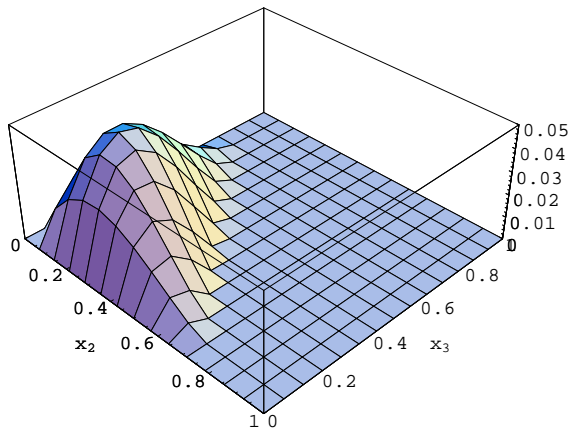,width=7.5cm}}
\caption[]{\sf Twist-3 distribution amplitude $\Phi_3(x_i)$}
\label{figure1} 
\end{figure}

\begin{figure}[h]
\centerline{
   \epsfig{file=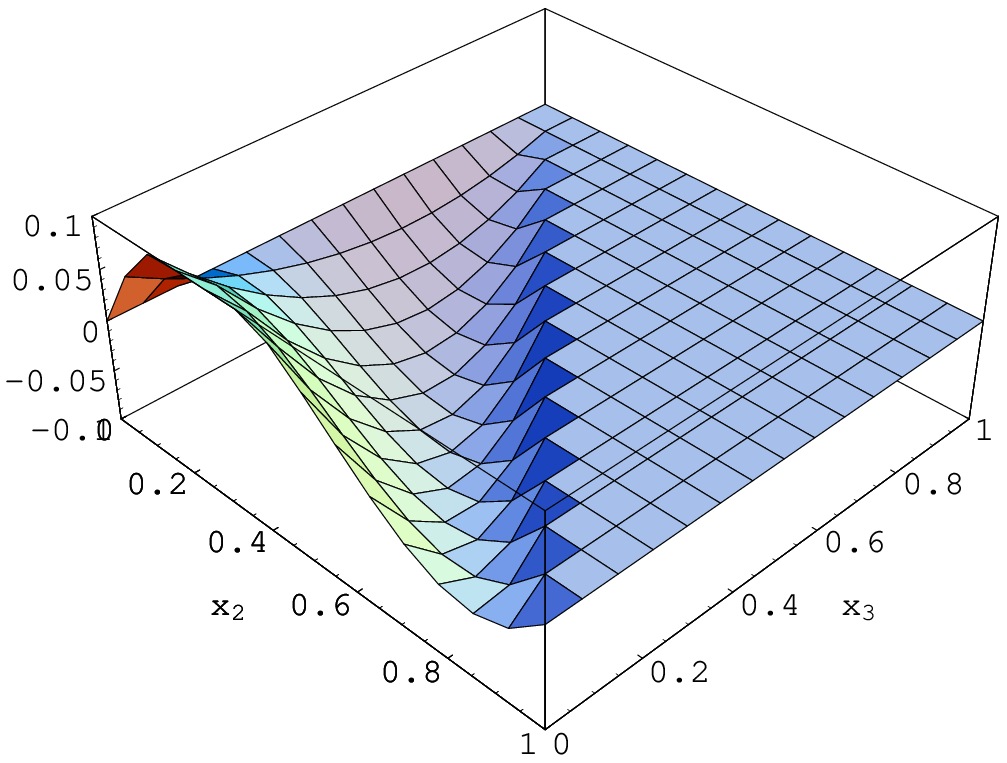,width=7.5cm}\hfill
   \epsfig{file=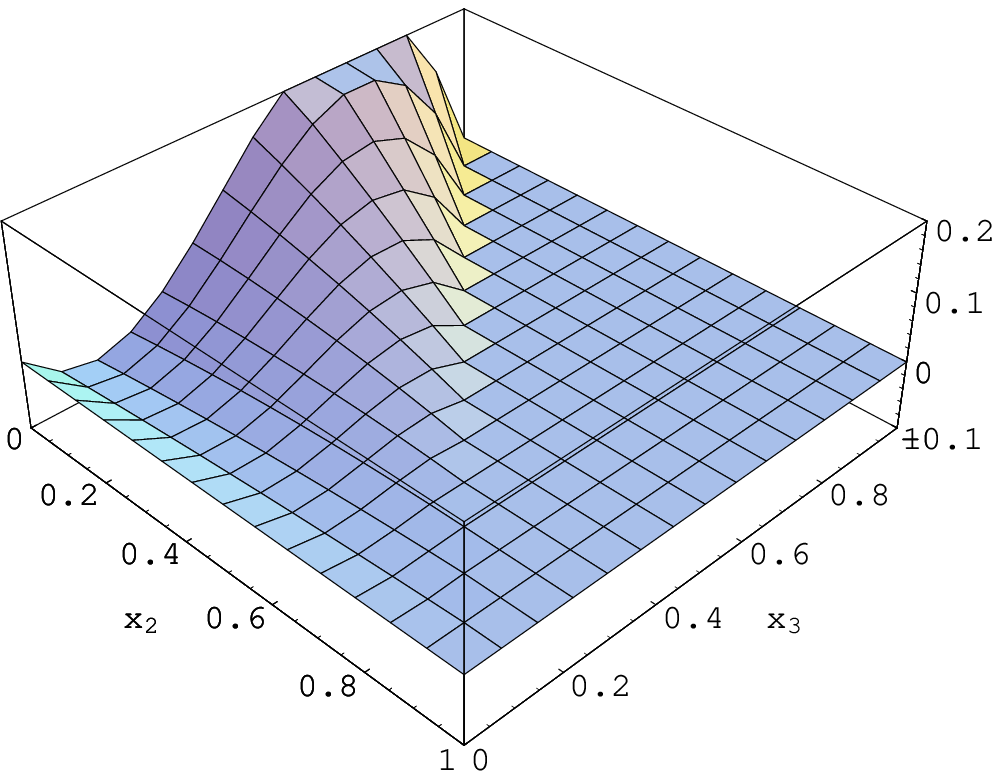,width=7.5cm}}
\centerline{
   \epsfig{file=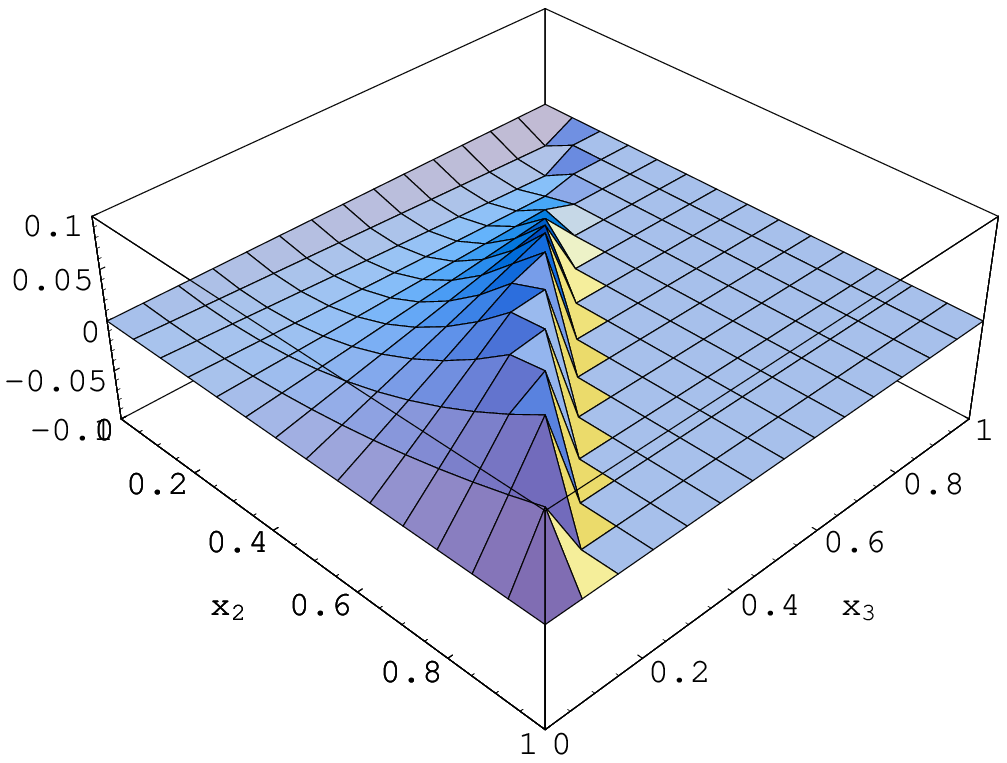,width=7.5cm}}
\caption[]{\sf Twist-4 distribution amplitudes $\Phi_4(x_i)$ in the first
line to the left, 
$\Psi_4(x_i)$ in the first line to the right,  
and $\Xi(x_i)$ in the second line}
\label{figure2} 
\end{figure}

\section{Summary and conclusions}

We have carried out a systematic study of the 
higher-twist light-cone distribution amplitudes of the
nucleon in QCD and found that a generic three-quark matrix
element on the light-cone can be parametrized in terms of 
eight independent nucleon distribution amplitudes. 
In particular, we have identified one distribution of leading twist-3, 
three of twist-4, three of twist-5 and one of twist-6. 
The shape of the distribution amplitudes at asymtotically large 
momentum transfers is found and is dictated by the conformal symmetry 
of the QCD Lagrangian. 
In order to quantify the corrections, we attempt to expand the 
distribution amplitudes in contributions of local conformal operators
and keep contributions of the next-to-leading conformal spin (``P''-waves).
The coefficients in this expansion present the necessary nonperturbative 
input. Some of them are related by QCD equations of motion, so that 
to our accuracy we end up with eight independent parameters 

From this amount, three parameters 
determine the leading twist amplitude
and are known already from the analysis in \cite{Che84}; two more
--- the normalisations of the higher-twist asymptotic distributions ---
can be directly inferred from QCD sum rule phenomenology of the nucleon.
The remaining three parameters  determine the deviation of the 
higher-twist distribution amplitudes from their asymptotic form   
and are estimated in this work.

To avoid confusion, we remind that our work does not 
present the complete analysis of all existing higher-twist distributions,
but of its subset related to three-quark operators without extra
gluon fields. As explained in the introduction, we believe that 
the three-quark contributions considered in this paper 
dominate higher-twist corrections because of large matrix elements. 
Indeed, we found that the matrix elements of higher-twist 
three-quark operators are of the same order or larger than those 
of the leading twist, see Sec.~3.   
We stress that the situation 
in the nucleon case is quite different to that of higher 
twist-corrections in the meson case studied earlier 
\cite{Bra90,BallBraun}. The reason is that in the meson
case ``bad'' components of quark fields always can be 
eliminated in favour of ``good'' components
and an additional gluon. The resulting  matrix elements turn out
to be numerical small. In the nucleon case quark fields
with ``minus'' projection give rise to  genuine three-quark 
higher twist effects that cannot be attributed 
to higher Fock components in the wave function, involving an additional gluon.

A detailed summary of the distribution amplitudes is presented 
in Appendix C. In this paper we do not consider phenomenological 
applications, but expect that our results are relevant for the 
studies of a wide range of interesting physical processes.
  
One obvious application can be found in 
the calculation of the nucleon form factors. While the magnetic
form factor of the proton $G_{\rm M}(Q^2)$ allows for 
the leading twist description \cite{exclusive,CZreport}, 
it is known that the Pauli formfactor $F_2(Q^2)$ 
is higher-twist, suppressed by an additional power of $Q^2$.
The description of $F_2$ therefore calls for the input 
of higher twist distribution amplitudes presented in this paper.

Twist-4 distribution amplitudes of the nucleon are, 
probably, the most interesting and in general are
related to various spin asymmetries in exclusive processes.
Indeed, while the total helicity is conserved 
in processes involving 
leading twist distribution amplitdues \cite{BL81}, 
this is not the case for non-leading twist contributions.
Thus spin-sensitive transition formfactors will provide  a 
testing ground for higher twist effects.

Last but not least, the distribution amplitudes presented in this 
paper can immediately be applied to calculations of 
exclusive reactions at moderate momentum 
transfers in the framework of QCD  light-cone sum rules 
\cite{LCsum}. This approach allows to study soft 
non-factorisable contributions to hard 
exclusive reactions in a largely model-independent way, see e.g. 
 \cite{LC2}.

\section*{Acknowledgements}
This work was supported by the DFG, project 920585 and Graduiertenkolleg
``Physik der starken Wechselwirkung''.

\appendix
\renewcommand{\theequation}{\Alph{section}.\arabic{equation}}
\setcounter{table}{0}
\renewcommand{\thetable}{\Alph{table}}

\section*{Appendices}

\section{Fierz transformations}
\label{app:a}
\setcounter{equation}{0}

In this section we give the Fierz transformation rules
necessary to derive the isospin constraints \Gl{isospin2} from 
the symmetry requirement \Gl{isospin}.

We write the definition 
of distribution amplitudes in \Gl{defdisamp} in a shorthand notation as 
\beqar{defdisamp2}
&& \bra{0} \ep^{ijk} u_\al^i(a_1 z) u_\be^j(a_2 z) d_\ga^k(a_3 z) 
\ket{P(P,\la)}
\nn \\
&& = 
S_1 (s_1)_{\al\be,\ga} + 
S_2 (s_2)_{\al\be,\ga} +  
P_1 (p_1)_{\al\be,\ga} +    
P_2 (p_2)_{\al\be,\ga} +  
V_1 (v_1)_{\al\be,\ga} +  
V_2  (v_2)_{\al\be,\ga} +  
\frac12 V_3 (v_3)_{\al\be,\ga} \nn \\ && +    
\frac12 V_4 (v_4)_{\al\be,\ga} +  
V_5 (v_5)_{\al\be,\ga} +  
V_6 (v_6)_{\al\be,\ga} +
A_1 (a_1)_{\al\be,\ga} +  
A_2 (a_2)_{\al\be,\ga} +  
\frac12 A_3 (a_3)_{\al\be,\ga} \nn \\ && +      
\frac12 A_4 (a_4)_{\al\be,\ga} +  
A_5 (a_5)_{\al\be,\ga} +  
A_6 (a_6)_{\al\be,\ga} +  
T_1 (t_1)_{\al\be,\ga} +  
T_2 (t_2)_{\al\be,\ga} +  
T_3 (t_3)_{\al\be,\ga} \nn \\ && +    
T_4 (t_4)_{\al\be,\ga} +  
T_5 (t_5)_{\al\be,\ga} +  
\frac12 T_7 (t_7)_{\al\be,\ga} +  
\frac12 T_8 (t_8)_{\al\be,\ga}    
\, .
\eeqar
The small letters, e.g.~$(s_1)_{\al\be,\ga} = \left(C\right)_{\al\be}
N_\ga$ stand for the Lorentz structures 
$\left(\Ga C\right)_{\al\be} \left(\Gamma' N \right)_\ga$, etc.
By means of the following Fierz transformation
\beqar{fierz}
&&\left(\Ga C\right)_{\al\be} \left(\Gamma' N \right)_\ga
= \frac14 \bigg[ C_{\ga\be} \left(\Ga\Ga'N\right)_\al + 
         \left(\ga_5 C\right) _{\ga\be} \left(\Ga\ga_5\Ga'N\right)_\al 
\nn \\ &&
+ \frac{1}{\pz}\left(\!\not\!{p} C\right) _{\ga\be} 
  \left(\Ga\!\not\!{z}\Ga'N\right)_\al 
+ \frac{1}{\pz}\left(\!\not\!{z} C\right) _{\ga\be} 
  \left(\Ga\!\not\!{p}\Ga'N\right)_\al 
+ \left(\ga_\perp C\right) _{\ga\be} 
  \left(\Ga\ga^\perp\Ga'N\right)_\al 
\nn \\ &&
- \frac{1}{\pz}\left(\!\not\!{p} \ga_5 C\right) _{\ga\be} 
  \left(\Ga\!\not\!{z}\ga_5 \Ga'N\right)_\al 
- \frac{1}{\pz}\left(\!\not\!{z} \ga_5 C\right) _{\ga\be} 
  \left(\Ga\!\not\!{p}\ga_5 \Ga'N\right)_\al 
- \left(\ga_\perp \ga_5 C\right) _{\ga\be} 
  \left(\Ga\ga^\perp\ga_5 \Ga'N\right)_\al 
\nn \\ &&
+ \frac{1}{(\pz)^2}\left(\si_{p z}C\right) _{\ga\be} 
  \left(\Ga\si_{z p}\Ga'N\right)_\al 
+ \frac{1}{2}\left(\si_{\perp\perp'}C\right) _{\ga\be} 
  \left(\Ga\si^{\perp\perp'}\Ga'N\right)_\al 
\nn \\ &&
+ \frac{1}{\pz}\left(i \si_{\perp p}C\right) _{\ga\be} 
  \left(\Ga\ga^\perp\!\not\!{z}\Ga'N\right)_\al 
+ \frac{1}{\pz}\left(i \si_{\perp z}C\right) _{\ga\be} 
  \left(\Ga\ga^\perp\!\not\!{p}\Ga'N\right)_\al \bigg]
\eeqar
all Lorentz structures are brought into the same form,
i.e.~we apply for the expansion of the third contribution 
in \Gl{isospin} the following transformations of twist-3 structures
\beqar{fierz-t3}
(v_1)_{\ga\be,\al} &=& \frac{1}{2}\left(  v_1 -  a_1 -   t_1
\right)_{\al\be, \ga} \nn \\
(a_1)_{\ga\be,\al} &=& \frac{1}{2} \left(-   v_1 +   a_1 -   t_1
\right)_{\al\be, \ga} \nn \\
(t_1)_{\ga\be,\al} &=& - \left(v_1 +  a_1 \right)_{\al\be,\ga}
\eeqar
and similar transformations for twist-4 
\beqar{fierz-t4}
(s_1)_{\ga\be,\al} &=& \frac{1}{4} \left(
s_1  + p_1- 2 v_2 - 2 a_2 + v_3 - a_3 - t_3 - 2 t_2 + \frac{1}{2} t_7
\right)_{\al\be, \ga}
\nn \\
(p_1)_{\ga\be,\al} &=& \frac{1}{4} 
\left(s_1  + p_1+ 2 v_2 + 2 a_2 - v_3 + a_3 - t_3 - 2 t_2 + \frac{1}{2} t_7\right)_{\al\be, \ga}
\nn \\
(v_2)_{\ga\be,\al} &=& \frac{1}{4} 
\left(-s_1  +  p_1  + v_3 +  a_3 - t_3 - \frac{1}{2} t_7
\right)_{\al\be, \ga} \nn \\
(a_2)_{\ga\be,\al} &=& \frac{1}{4} 
\left(-s_1  +  p_1  - v_3 -  a_3 - t_3 - \frac{1}{2} t_7
\right)_{\al\be, \ga} \nn \\
(v_3)_{\ga\be,\al} &=& \frac{1}{4} 
\left(2 s_1  - 2 p_1  + 4 v_2 - 4 a_2 - 2 t_3 - t_7
\right)_{\al\be, \ga} \nn \\
(a_3)_{\ga\be,\al} &=& \frac{1}{4} 
\left(- 2 s_1  + 2 p_1  + 4 v_2 - 4 a_2 + 2 t_3 + t_7
\right)_{\al\be, \ga}
\nn \\
(t_2)_{\ga\be,\al} &=& \frac{1}{4} \left(
- 2 s_1 - 2 p_1 - 2 t_3 + t_7\right)_{\al\be, \ga}
\nn \\
(t_3)_{\ga\be,\al} &=& \frac{1}{4} \left(
- s_1 - p_1 - 2 v_2 -2 a_2 - v_3 + a_3 + t_3 - 2 t_2 - \frac{1}{2} t_7
\right)_{\al\be, \ga}
\nn \\
(t_7)_{\ga\be,\al} &=& \frac{1}{4} \left(
2 s_1 + 2  p_1 - 4 v_2 -4 a_2 - 2 v_3 + 2 a_3 - 2  t_3 + 4 t_2 + t_7
\right)_{\al\be, \ga}
\, .
\eeqar
Relations for twist-5 and twist-6 are identical
up to the substitutions: $\{v_1,a_1,t_1\} \to \{v_6,a_6,t_6\}$
and $\{s_1,p_1,v_2,v_3,a_2,a_3,t_2,t_3,t_7\} \to 
\{s_2,p_2,v_5,v_4,a_5,a_4,t_5,t_4,t_8\}$.
By a simple substitution, one also gets the necessary  
Fierz transformations from $(\ldots)_{\al\ga,\be} \to 
(\ldots)_{\al\be,\ga}$ required in the second contribution in \Gl{isospin}.
Putting everything together we derive three equations for
the twist-3 amplitudes from the coefficients $v_1,a_1,t_1$:
\beqar{ausfuer}
&& 
0 = \left[2 T_1(1,2,3) - V_1(1,3,2) + A_1(1,3,2) - V_1(3,2,1) -
A_1(3,2,1)\right]
\left(t_1\right)_{\al\be, \ga}
\nn \\ && + 
\left[2 V_1(1,2,3) + 2 V_1(1,3,2) + 2 A_1(1,3,2) - 2 T_1(1,3,2) 
\right.\nn \\ &&  \left. \quad +
V_1(3,2,1) - A_1(3,2,1)- 2 T_1(3,2,1) \right]
\left(v_1\right)_{\al\be, \ga}
\nn \\ && + 
\left[2 A_1(1,2,3) + 2 V_1(1,3,2) + 2 A_1(1,3,2) + 2 T_1(1,3,2)  
\right.\nn \\ &&   \left. \quad
- V_1(3,2,1) + A_1(3,2,1)- 2 T_1(3,2,1) \right]
\left(a_1\right)_{\al\be, \ga} \, .
\eeqar
Using the symmetry properties \Gl{symm} it is easy to see,
however,  that all three above equations are in fact identical 
to the one given in \Gl{isospin2}.  
In the similar way one can derive an overcomplete set of nine 
twist-4 equations that all can be reduced to the two equations 
relating twist-4 amplitudes in \Gl{isospin2}. 
The derivation for the remaining twist-5 and twist-6
equations is similar.

\section{Nonperturbative parameters}

\label{app:b}
\setcounter{equation}{0}

In Sect.~3.2  we have demonstrated that the normalisation of the asymptotic
distribution amplitudes of all twists involves three 
independent parameters.
In the first part of this  Appendix we consider leading corrections to 
the asymptotic distribution amplitudes, related to contributions of 
three-quark conformal one extra derivative, and find that they involve  
five new parameters.

In the second part, we work out QCD sum rule estimates of the higher-twist
matrix elements. 
Partially the results can be adapted from existing
sum rule calculations containing nucleon currents 
\cite{Iof81,Kol84, Lein94}, partially they are new.
\subsection{Equations of motion}
The coefficients of ``P''-wave contributions in the conformal expansion 
of distribution amplitudes are given by matrix elements of operators
containing one derivative. In case the derivative acts on the 
down quark we have the general decomposition:
\beqar{zerld}
&& 4 \bra{0} \ep^{ijk} u_\al^i(0) u_\be^j(0) 
\left[ iD_\la d_\ga\right]^k(0) \ket{N,P}
\nn \\
&& = 
{\cal V}_1^d P_\la  \left(\!\not\!{P}C \right)_{\al \be} \left(\ga_5 N\right)_\ga 
+ \V_2^0 M \left(\!\not\!{P} C \right)_{\al \be} \left(\ga_\la \ga_5 N\right)_\ga 
+ 
\V_3^d P_\la  M  \left(\ga_\mu C \right)_{\al \be}\left( \ga^{\mu} 
\ga_5 N\right)_\ga 
\nn \\ 
&& +
\V_4^0 M^2 \left(\ga_\la C \right)_{\al \be} \left(\ga_5 N\right)_\ga +
\V_5^0 M^2 \left(\ga^\mu C \right)_{\al \be} \left(i \si_{\mu\la} \ga_5 
N\right)_\ga 
\nn \\ 
&& +
\T_1^d P_\la \left(P^\nu i \si_{\mu\nu} C\right)_{\al \be} 
\left(\ga^\mu\ga_5 N\right)_\ga + 
\T_2^0 M \left(P^\nu i \si_{\la\nu} C\right)_{\al \be} 
\left(\ga_5 N\right)_\ga +
\T_3^d P_\la M \left(\si_{\mu\nu} C\right)_{\al \be} 
\left(\si^{\mu\nu}\ga_5 N\right)_\ga 
\nn \\
&& 
+ 
\T_4^0 M \left(P_\nu \si^{\mu\nu} C\right)_{\al \be} 
\left(\si_{\mu\la} 
\ga_5 N\right)_\ga 
+ 
\T_5^0 M^2 \left(i \si_{\mu\la} C\right)_{\al \be} 
\left(\ga^\mu\ga_5 N\right)_\ga 
\nn \\
&& +
\T_{7}^0 M^2 \left(\si_{\mu\nu} C\right)_{\al \be} 
\left(\si^{\mu\nu} \ga_\la \ga_5 N\right)_\ga
\eeqar
yielding 11 parameters.

Acting with one derivative on the up-quarks picks up  
the structures of  $\ga$-matrices that are odd under 
transposition:
\beqar{zerlu}
&& 4 \bra{0} \ep^{ijk} u_\al^i(0) i \Dlr_\la
u_\be^j(0) d_\ga^k(0) \ket{N,P}
\nn \\
&& = 
{\cal S}_1^u P_\la  M C_{\al \be} \left(\ga_5 N\right)_\ga + 
{\cal S}_2^0 M^2 C_{\al \be} \left(\ga_\la \ga_5 N\right)_\ga + 
\nn \\
&& + 
{\cal P}_1^u P_\la M \left(\ga_5 C\right)_{\al \be} N_\ga + 
{\cal P}_2^0 M^2 \left(\ga_5 C \right)_{\al \be} \left(\ga_\la N\right)_\ga + 
\nn \\
&& + 
\A_1^u P_\la \left(\!\not\!{P}\ga_5 C \right)_{\al \be} N_\ga + 
\A_2^0 M \left(\!\not\!{P}\ga_5 C \right)_{\al \be} \left(\ga_\la N\right)_\ga  + 
\A_3^u P_\la M  \left(\ga_\mu \ga_5 C \right)_{\al \be}
\left( \ga^{\mu} N\right)_\ga 
\nn \\ 
&& +
\A_4^0 M^2 \left(\ga_\la \ga_5 C \right)_{\al \be} N_\ga +
\A_5^0 M^2 \left(\ga^\mu \ga_5 C \right)^{\al \be} \left(i \si_{\mu\la}
N\right)_\ga 
\eeqar
and introduces 9 parameters. Hence, altogether there are 20 unknown numbers.
The equations of motions yield, however,  a number of constraints:
\beqar{constraint}
&& 
\bra{0} \ep^{ijk} u^i(0) C \ga_\ro u^j(0) \,
\ga^\la \left[iD_\la d_\ga\right]^k(0) \ket{N,P} = 0 \,,
\nn \\
&&
\bra{0} \ep^{ijk} u^i(0) C \ga^\la u^j(0) \,
\left[iD_\la d_\ga\right]^k(0) \ket{N,P} = 
P_\la
\bra{0} \ep^{ijk} u^i(0) C \ga_\la u^j(0) \,
d_\ga^k(0) \ket{N,P} \,,
\nn \\
&&
\bra{0} \ep^{ijk} u^i(0) C \si_{\al\be}  u^j(0) \,
\ga^\la \left[iD_\la d_\ga\right]^k(0) \ket{N,P} = 0 \,,
\nn \\ && 
\bra{0} \ep^{ijk} u^i(0) C i \si_{\al\be}  u^j(0) \,
\left[iD^\be d_\ga\right]^k(0) \ket{N,P} \nn \\ &&
= P^\be \bra{0} \ep^{ijk} u^i(0) C i \si_{\al\be}  u^j(0) 
\, d^k(0) \ket{N,P} 
- 
\bra{0} \ep^{ijk} \left[u(0) C i \Dlr_\al  u(0)\right]^{ij} \, 
d_\ga^k(0) \ket{N,P} \,,
\nn \\ &&
\bra{0} \ep^{ijk} u^i(0) C i \ga_5\si_{\al\be}  u^j(0) \,
\, \left[iD^\be d_\ga\right]^k(0) \ket{N,P} \nn \\ &&
= P^\be \bra{0} \ep^{ijk} u^i(0) C \ga_5 i \si_{\al\be} u^j(0) \,
d^k(0) \ket{N,P} 
- 
\bra{0} \ep^{ijk} \left[u(0) C \ga_5 i \Dlr_\al  u(0) \right]^{ij}  
\, d_\ga^k(0) \ket{N,P} \,,
\nn \\
\eeqar 
which translate to 
\beqar{impl1}
\V_1^d &=& 4 \V_2^0 + 2 \V_3^d 
\,,\hspace{0.8cm}\qquad 
- 3 \V_5^0 = \V_3^d +  \V_4^0 \,,
\nn \\ 
\V_1^0- \V_3^0  &=& \V_1^d - \V_3^d + 4 \V_4^0 - \V_2^0 \,, 
\nn \\ 
0 &=& \T_3^d + \T_5^0  
\,,\hspace{1.6cm} \qquad  
\T_2^0 = - \T_1^d + 4 \T_3^d + 3 \T_4^0 \,,
\nn \\ 
\T_1^0 - 2 \T_3^0 - {\cal S}_2^0 &=&
\T_1^d - 2 \T_3^d - \T_4^0 + 3 \T_5^0 + 6 \T_7^0 
\,,
\nn \\ 
\T_1^0 - 2 \T_3^0 - {\cal S}_1^u &=&
\T_1^d - 3 \T_2^0 - 2 \T_3^d - \T_4^0  
\,,
\nn \\ 
- 2 \T_3^d + 2 \T_4^0 &=& 
- 2 \T_3^0 - {\cal P}_1^u 
\,,\hspace{0.3cm}\quad 
2 \T_3^0 - {\cal P}_2^0 = 
2 \T_3^d - 2 \T_4^0 - 6 \T_7^0 
\,,
\eeqar
The two relations in the first line in \Gl{impl1} 
follow from the first equation in \Gl{constraint}, the relations in the 
second line from the second equation, the third line from the 
third equation, fourth and fifth line from the fourth equation and
the last line from the last equation.
In a similar way one observes that the axial structures 
have to fulfill the constraints
\beqar{constraint2}
&& 
\bra{0} \ep^{ijk} \left[u(0) C \ga^\ro \ga_5 \Dlr_\ro u(0) \right]^{ij}\, 
d_\ga^k(0) \ket{N,P} = 0
\,, 
\nn \\
&&\bra{0} \ep^{ijk} \left[u(0) C \left\{\ga_\la i\Dlr_\ro -
\ga_\ro i\Dlr_\la\right\} \ga_5 u(0)\right]^{ij} 
d_\ga^k(0) \ket{N,P} \nn 
\\ && 
= 
\bra{0} \ep^{ijk} \left[u(0) C 
\frac{i}{2}\left\{\si_{\la\ro} \ga^\al i\vecl{D}_\al +
\ga^\al \si_{\la\ro} i\vecr{D}_\al\right\} \ga_5 u(0) \right]^{ij}\,
d_\ga^k(0) \ket{N,P} \nn \\ &&
=  
- i \ep_{\la\ro\al\de} \left[
P^\al
\bra{0} \ep^{ijk} u^i(0) C \ga^\de u^j(0) d_\ga^k(0) \ket{N,P} 
-
\bra{0} \ep^{ijk} u^i(0) C \ga^\de u^j(0) 
\left[i D^\al d_\ga\right]^k(0) \ket{N,P}\right] \,,
\nn \\
\eeqar
where from it follows that
\beqar{impl3}
&& \A_1^u + \A_3^u + \A_2^0 + 4 \A_4^0 =0 \,,
\nn \\
&& \A_3^u - \A_2^0 = \V_2^0 + \V_3^0 - \V_3^d \,, 
\qquad
2 \A_5^0 = \V_2^0 + \V_3^0 - 2 \V_5^0 - \V_3^d \,.
\eeqar
where the relation in the first line is derived from the first equation in 
\Gl{constraint2}.
Combining the constraints from the equations of motion with 
the eight isospin relations obtainable from \Gl{isospin2} 
we obtain an overcomplete
set of 20 equations. Choosing 
$\V_1^d,\A_1^u,\V_3^d,\A_3^u$
and $\T_3^d$ as the independent parameters and solving these equations, 
we can express the remaining parameters as
\beqar{supereqn}
&& \V_2^0 = 
   \frac{1}{4}\left(\V_1^d - 2\,\V_3^d\right) \,, \qquad
\hspace{4.2cm}
\V_4^0 = \frac{1}{16}\left(
        4\,\V_1^0  - 3\,\V_1^d - 
        4\,\V_3^0 + 2\,\V_3^d \right) \,, 
\nn \\ &&
\V_5^0 = \frac{1}{48}\left(
        -4\,\V_1^0  + 3\,\V_1^d + 
        4\,\V_3^0 - 18\,\V_3^d\right)
\,, \qquad  \hspace{0.9cm}
\A_2^0 = 
     \A_3^u - \frac{\V_1^d}{4} - \V_3^0 + 
      \frac{3\,\V_3^d}{2}
\,, 
\nn \\ &&
\A_4^0= \frac{1}{16}\left(
        -4\,\A_1^u - 8\,\A_3^u + \V_1^d + 
        4\,\V_3^0 - 6\,\V_3^d\right)
\,, \qquad
\A_5^0 = \frac{1}{48}\left(
        4\,\V_1^0 + 3\,\V_1^d + 
        20\,\V_3^0 - 18\,\V_3^d\right)
\,, 
\nn \\ &&
\T_2^0  = \frac{1}{8}\left(
         \A_1^u + 6\,\A_3^u + 8\,\T_3^d - 
        \V_1^0 - 2\,\V_1^d - 6\,\V_3^0 + 
        12\,\V_3^d\right)
\,, 
\qquad
\T_5^0 = -\T_3^d
\,, \nn \\ &&
\T_4^0 = \frac{1}{8}\left(
         -\A_1^u + 2\,\A_3^u - 8\,\T_3^d + 
        \V_1^0 - 2\,\V_1^d - 2\,\V_3^0 + 
        4\,\V_3^d \right)
\,, 
\nn \\ &&
\T_7^0 = \frac{1}{96}\left(
        3\,\A_1^u - 6\,\A_3^u - 
        16\,\T_3^0 + 48\,\T_3^d + \V_1^0 + 
        6\,\V_1^d + 6\,\V_3^0 - 12\,\V_3^d\right)
\,, \nn \\ &&
{\cal S}_2^0  = \frac{1}{16}\left(
        3\,\A_1^u + 10\,\A_3^u - 
        16\,\T_3^0 + 16\,\T_3^d + 9\,\V_1^0 - 
        2\,\V_1^d - 10\,\V_3^0 + 20\,\V_3^d\right) \,,
\nn \\ &&
{\cal S}_1^u  = \frac{1}{4}\left(
        3\,\A_1^u + 10\,\A_3^u - 
        8\,\T_3^0 + 16\,\T_3^d + \V_1^0 - 
        2\,\V_1^d - 10\,\V_3^0 + 20\,\V_3^d\right) \,,
\nn \\ &&
{\cal P}_1^u  = \frac{1}{4}\left(
         \A_1^u - 2\,\A_3^u - 8\,\T_3^0 + 
        16\,\T_3^d - \V_1^0 + 2\,\V_1^d + 
        2\,\V_3^0 - 4\,\V_3^d\right) \,,
\nn \\ &&
{\cal P}_2^0 = \frac{1}{16}\left(
        -\A_1^u + 2\,\A_3^u + 16\,\T_3^0 - 
        16\,\T_3^d + 5\,\V_1^0 - 2\,\V_1^d - 
        2\,\V_3^0 + 4\,\V_3^d\right)\,,
\nn \\ &&
\T_1^d = \frac{1}{2}\left(
     -\A_1^u + \V_1^0 - \V_1^d\right)  \,.
\eeqar
Finally, the parameters $\V_1^d,\A_1^u,\V_3^d,\A_3^u$ and $\T_3^d$ 
can be expressed by the reduced matrix elements of the
operators defined in \Gl{local-operators-2}. We have
\beqar{vergleich}
\V_1^d &=& f_N V_1^d\,, \qquad
\A_1^u = f_N A_1^u\,, \qquad
\la_1 f_1^d = \frac{1}{2}\left(\V_1^d - 6 \V_3^d\right) \,,
\nn \\
\la_1 f_1^u &=& \frac{1}{2}\left( - 2 \A_1^u - 12 \A_3^u + \V_1^d + 4
                               \V_3^0 - 6 \V_3^d\right)\,,
\nn \\ 
\la_2 f_2^d &=& 2 \left( - 2 \A_3^u - 8 \T_3^d + \V_1^d +2 \V_3^0
- 4 \V_3^d + 2 \T_1^d\right) \,.
\eeqar
The first two parameters appear already at the level of twist-3 and 
are known from \cite{Che84}. The others will be calculated
in the next subsection.

\subsection{QCD sum rule estimates}



The QCD sum rule estimates for $\la_1$, $\la_2$ are derived from the 
consideration of the two-point correlation function 
\beqar{kolesni}
i \int \dd^4 x\, e^{i q x}
\bra{0} \eta_{i}(x) \bar{\eta}_{i}(0) \ket{0}
= |\la_i|^2 M^2\frac{(\!\not\!{q} + M)}{M^2 - q^2}
+ \ldots\,,
\eeqar
where $\bra{0}\eta_1(0)\ket{P} = M \la_1 N(P)$ and
$\bra{0}\eta_2(0)\ket{P} = M \la_2 N(P)$\footnote{Note that our
normalisation of $\lambda_i$ differs by a factor $M$ from the standard 
one} are the local three-quark operators 
defined in \Gl{local-operators}. The dots refer to excited states
and the continuum contribution.
Taking into account vacuum condensates up to dimension-8 
one obtains the sum rules \cite{Kol84,Lein94}:
\beqar{sumrule1}
2 (2\pi)^4 M^2 |\la_1|^2 &=& \exp(M^2/M_B^2) \left\{ 
M_B^6 E_3(s_0/M_B^2)
+ \frac{b}{4} 
M_B^2 E_1(s_0/M_B^2) 
+ \frac{a^2}{3} \left(4 - \frac{m_0^2}{M_B^2}\right)\right\}
\nn \\
\eeqar
and
\beqar{sumrule2}
2 (2\pi)^4 M^2\frac{|\la_2|^2}{6} &=& \exp(M^2/M_B^2) \left\{ M_B^6
E_3(s_0/M_B^2)
+ \frac{b}{4} M_B^2 E_1(s_0/M_B^2) 
\right\}
\,,
\eeqar
where 
\beqar{continuum} 
E_n(s_0/M_B^2) = 1 - e^{(-s_0/M_B^2)} \sum_{k=0}^{n-1} 
\frac{1}{k!} \left(\frac{s_0}{M_B^2}\right)^k \,.
\eeqar
Here and below $M_B$ is the Borel parameter.
We use  the Borel window $1 {\rm GeV^2} \leq M_B^2 \leq 2 {\rm GeV^2}$, 
with the continuum threshold $\sqrt{s_0} \sim 1.5 \;{\rm GeV}$ and 
values of the condensates normalized at $\mu^2 = 1~{\rm GeV^2}$ 
\cite{BallBraun}:
\beqar{condenstates}
a &=& - (2\pi)^2 \langle \bar q q \rangle 
~~~~~\sim 0.55 \; {\rm
GeV^3} \,,
\nn \\
b &=& \phantom{-}(2\pi)^2 \langle \frac{\al_S}{\pi} G^2\rangle 
\sim 0.47 \; {\rm
GeV^4} \,,
\nn \\
m_0^2  &=& \phantom{-}\frac{\langle \bar q g G q \rangle}{\langle \bar q
q\rangle} 
~~~~~~~~\sim 0.8 \; {\rm GeV^2} \,.
\eeqar
With these inputs, 
we find $|\la_1| = 0.027\pm 0.009\, {\rm GeV^2} $, $|\la_2| = 0.051\pm
0.019 \, {\rm GeV^2}$. Note that the above sum rules only fix the 
absolute value of the parameters $\la_1$ and $\la_2$.
Relative phases between different nucleon-to-vacuum matrix elements  
can be computed by investigating
suitable non-diagonal correlation functions.
To determine the relative sign between $f_N$ and $\la_1$ we consider
the correlation function
\beqar{sign}
i \int \dd^4 x\, e^{i q x}
\bra{0}  \ep^{ijk} \left[u^i(x) C \!\not\!{z} u^j(x)\right] \,
\ga_5 \!\not\!{z} d^k(x)\; \bar{\eta}_{1}(0) \ket{0}
= f_N \la_1^* M \!\not\!{z}\frac{(\!\not\!{q} + M)}{M^2 - q^2}
+ \ldots \,.
\eeqar
Taking the ratio of the corresponding sum rule and the  
sum rule \Gl{sumrule1} we obtain
\beqar{sign-sumrule}
\frac{f_N}{\lambda_1} = - \frac{1}{3} 
\frac{\left(2 M_B^2 E_1(s_0/M_B^2) - m_0^2\right) a M}{\displaystyle
M_B^6 E_3(s_0/M_B^2)
+ \frac{b}{4} 
M_B^2 E_1(s_0/M_B^2) 
+ \frac{a^2}{3} \left(4 - \frac{m_0^2}{M_B^2}\right)} 
\eeqar
which is real and negative. 
We have chosen $f_N$ to be positive according
to the standard choice in \cite{Che84} and $\la_1$ to be negative.
Note also that the above sum rule leads to a value 
$f_N = 5.6 \cdot 10^{-3} {\rm GeV^2}$ consistent with standard estimates.
A similar calculation of the nondiagonal correlation function 
\beqar{kol2}
i \int \dd^4 x\, e^{i q x} \bra{0}
\eta_{1}(x) \bar{\eta}_{2}(0)  + 
\eta_{2}(x) \bar{\eta}_{1}(0) \ket{0}
= \left(\la_1 \la_2^* + \la_2 \la_1^*\right) M^2
\frac{(\!\not\!{q} + M)}{M^2 - q^2}
+ \ldots 
\eeqar
was performed in \cite{Kol84}. The resulting 
sum rule
\beqar{sumrulekolya} 
- 2 (2\pi)^4 M^3 \left(\la_1 \la_2^* + \la_2 \la_1^*\right) 
= 2 a e^{\frac{M^2}{M_B^2}} \left(12 M_B^4 E_2(s_0/M_B^2) 
- 6 m_0^2 M_B^2 E_1(s_0/M_B^2) + \frac{32 \al_s a^2}{27 \pi M_B^2}\right)
\nn \\
\eeqar
shows that the relative sign between $\la_2$ and $\la_1$ is negative.

To estimate the remaining parameters we consider the nondiagonal
correlation functions  involving the last three higher-twist operators
in \Gl{local-operators-2} with either 
$\bar\eta_1(0)$ or $\bar\eta_2(0)$:
\beqar{cor1}
&&i \int \dd^4 x \, e^{i q x}
\bra{0} \ep^{ijk} \left(u^i(x) C \ga_\mu u^j(x)\right) \,
\ga_5 \ga^\mu (i z \vecr{D} d)^k(x) \;
\bar{\eta}_{1}(0) \ket{0} 
= \frac{f_1^d |\la_1|^2 M^2 q\cdot z \!\not\!{q}}{M^2 - q^2}
+ \ldots
\,, \nn \\
&&i \int \dd^4 x \, e^{i q x} 
\bra{0} \ep^{ijk} 
\left(u(x) C \ga_\mu \ga_5 i z\Dlr u(x)\right)^{ij} \,
\ga^\mu d^k(x) \;
\bar{\eta}_{1}(0) \ket{0}
= \frac{f_1^u |\la_1|^2 M^2 q\cdot z \!\not\!{q}}{M^2 - q^2}
+ \ldots
\,, \nn \\
&& i \int \dd^4 x \, e^{i q x}
\ep^{ijk} \bra{0}\left(u^i(x) C\si_{\mu\nu} u^j(x)\right) 
\,\ga_5 \si^{\mu\nu} (i z \vecr{D} d)^k(x) \;
\bar{\eta}_{2}(0) \ket{0} 
= \frac{f_2^d |\la_2|^2 M^2 q\cdot z \!\not\!{q}}{M^2 - q^2}
+ \ldots
\,.
\nn \\
\eeqar
The dots refer, again,  to contributions of excited states 
and different Lorentz structures that we do not consider.
Expressing $|\la_i|^2$ by the expression in \Gl{sumrule1} and
\Gl{sumrule2} we obtain the following sum rules:
\beqar{oursumrules}
f_1^d &=&  
\frac{\displaystyle \frac{3}{10} M_B^6 E_3(s_0/M_B^2) 
+ \frac{b}{9} M_B^2 E_1(s_0/M_B^2) 
+ \frac{a^2}{3} \left(4 - \frac{m_0^2}{M_B^2}\right)
}{\displaystyle M_B^6 E_3(s_0/M_B^2) + \frac{b}{4} M_B^2 E_1(s_0/M_B^2) 
+ \frac{a^2}{3} \left(4 - \frac{m_0^2}{M_B^2}\right)} \,,
\nn \\
f_1^u &=&  
\frac{\displaystyle \frac{1}{10} M_B^6 E_3(s_0/M_B^2) 
+ \frac{b}{8} M_B^2 E_1(s_0/M_B^2) 
+ \frac{5 a^2}{18} \frac{m_0^2}{M_B^2}
}{\displaystyle M_B^6 E_3(s_0/M_B^2) + \frac{b}{4} M_B^2 E_1(s_0/M_B^2) 
+ \frac{a^2}{3} \left(4 - \frac{m_0^2}{M_B^2}\right)} 
\, ,
\nn \\
f_2^d &=&  
\frac{\displaystyle \frac{8}{5} M_B^6 E_3(s_0/M_B^2) -
\frac{4 b}{9} M_B^2 E_1(s_0/M_B^2) 
}{\displaystyle 
6 M_B^6 E_3(s_0/M_B^2) + \frac{3 b}{2} M_B^2 E_1(s_0/M_B^2)} 
\, .
\eeqar
Inserting the numerical values of parameters, we end up with 
the following estimates for the higher-twist matrix elements:
\beqar
{numericalvalues}
  f_1^d(\mu = 1~{\rm GeV}) &=&  0.6\pm 0.2  \,,
\nn\\
  f_2^d(\mu = 1~{\rm GeV}) &=&  0.15\pm 0.06 \,,  
\nn\\
  f_1^u(\mu = 1~{\rm GeV}) &=&  0.22\pm 0.15 \,.
\eeqar

\section{Handbook of nucleon distribution amplitudes}
\label{app:c}
\setcounter{equation}{0}

In this Appendix we give a complete list of all 24 nucleon distribution
amplitudes. 

The chiral structure of our basic set of eight independent 
distribution amplitudes is summarized in Table~2 in the text. 
The corresponding representations for the remaining 16 distributions
are collected in Table~\ref{tabellea} and  Table~\ref{tabelleb}.
The set in
Table~\ref{tabellea} contains the amplitudes that can be directly obtained
from \Gl{vector-twist-3} to \Gl{vector-twist-6} by 
flipping spin-projections of the two up-quarks.
Table~\ref{tabelleb} contains four distributions that are obtainable 
from the previous ones by exchanging light-cone projections, and also 
$T_1, T_2, T_5$  and $T_6$ that cannot be read off 
from the previous expressions in a straightforward
way and therefore are given below:
\beqar{tensor-twist-3}
\hspace*{-1cm}
\bra{0} \ep^{ijk}\! 
\left(u^{\up}_i(a_1 z) C i \si_{\perp z} u^{\up}_j(a_2 z)\right)  
\ga^\perp\!\not\!{z} d^{\down}_k(a_3 z) \ket{P} 
&=&  - 2 pz \!\not\!{z} N^\up\! \!\int\!\! \D x 
\,e^{-i pz \sum x_i a_i}\, 
T_1(x_i)\,,
\nn \\
\bra{0} \ep^{ijk}\! 
\left(u^{\down}_i(a_1 z) C i \si_{\perp z} u^{\down}_j(a_2 z)\right)  
\ga^\perp\!\not\!{p} d^{\down}_k(a_3 z) \ket{P} 
&=&  - 2 pz \!\not\!{p} N^\up\! \!\int\!\! \D x 
\,e^{-i pz \sum x_i a_i}\, 
T_2(x_i)\,,
\nn \\
\bra{0} \ep^{ijk}\! 
\left(u^{\down}_i(a_1 z) C i \si_{\perp p} u^{\down}_j(a_2 z)\right)  
\ga^\perp\!\not\!{z} d^{\down}_k(a_3 z) \ket{P} 
&=&  - M^2 \!\not\!{z} N^\up\! \!\int\!\! \D x 
\,e^{-i pz \sum x_i a_i}\, 
T_5(x_i)\,,
\nn \\
\bra{0} \ep^{ijk}\! 
\left(u^{\up}_i(a_1 z) C i \si_{\perp p} u^{\up}_j(a_2 z)\right)  
\ga^\perp\!\not\!{p} d^{\down}_k(a_3 z) \ket{P} 
&=&  - M^2\!\not\!{p} N^\up\! \!\int\!\! \D x 
\,e^{-i pz \sum x_i a_i}\, 
T_6(x_i)\,.
\eeqar
Note also that for scalar and tensor contributions the following 
representations can be given:
\beqar{additional}
\hspace*{-1cm}
\bra{0} \ep^{ijk}\! 
\left(u^{\up}_i(a_1 z) C u^{\up}_j(a_2 z)\right)  
\!\not\!{z} d^{\up}_k(a_3 z) \ket{P} 
&=&  \frac12 M \!\not\!{z} N^\up\! \!\int\!\! \D x 
\,e^{-i pz \sum x_i a_i}\, 
\left[S_1+P_1\right](x_i)\,,
\nn \\
\bra{0} \ep^{ijk}\! 
\left(u^{\up}_i(a_1 z) C i \si_{z p}u^{\up}_j(a_2 z)\right)  
\!\not\!{z} d^{\up}_k(a_3 z) \ket{P} 
&=&  \frac12 pz M \!\not\!{z} N^\up\! \!\int\!\! \D x 
\,e^{-i pz \sum x_i a_i}\, 
\left[T_3-  T_7\right](x_i)\,.
\nn \\
\eeqar
\begin{table}
\renewcommand{\arraystretch}{1.3}
\begin{center}
\begin{tabular}{l|l|l|l}
&  
Lorentz-structure 
&  
Light-cone projection   
& 
nomenclature
\\ \hline
twist-3   
&  $ \left(C\!\not\!{z}\right) \otimes \!\not\!{z} $    
& $u^+_\down u^+_\up d^+_\up$
& $\left[V_1+A_1\right](x_i)$ \\ \hline
twist-4
&  $ \left(C\!\not\!{z}\right) \otimes \!\not\!{p} $    
& $u^+_\down u^+_\up d^-_\up$
& $\left[V_2+A_2\right](x_i)$ 
\\ \hline
& $ \left(C\!\!\not\!{z}\ga_\perp\!\!\not\!{p}\,\right) 
      \otimes \ga^\perp\!\!\not\!{z} $    
& $ u^+_\down u^-_\up d^+_\down$
& $\left[V_3+A_3\right](x_i)$ \\ \hline
& $ \left(C \!\not\!{p}\!\not\!{z}\right)  \otimes \!\not\!{z}$    
& $u^-_\down u^+_\down  d^+_\up$
& $
\left[T_3 +  T_7 + S_1 - P_1 \right](x_i)$ \\ \hline
twist-5
&  $ \left(C\!\not\!{p}\right) \otimes \!\not\!{z} $    
& $u^-_\down u^-_\up d^+_\up$
& $\left[V_5+A_5\right](x_i)$ 
\\ \hline
& $ \left(C\!\!\not\!{p}\ga_\perp\!\!\not\!{z}\,\right) 
\otimes \ga^\perp\!\!\not\!{p} $    
& $ u^-_\down u^+_\up d^-_\down
$
& $\left[V_4+A_4\right](x_i)$ \\ \hline
& $ \left(C \!\not\!{z}\!\not\!{p}\right)  \otimes \!\not\!{p}$    
& $u^+_\down u^-_\down  d^-_\up$
& $ 
\left[T_4 +  T_8 + S_2 - P_2 \right](x_i)$ \\ \hline
twist-6   
&  $ \left(C\!\not\!{p}\right) \otimes \!\not\!{p} $    
& $u^-_\down u^-_\up d^-_\up$
& $\left[V_6+A_6\right](x_i)$ \\ \hline
\end{tabular}
\end{center} 
\caption[]{\sf 
 Eight nucleon distribution amplitudes 
that can be obtained from \Ta{tabelle1} by flipping the spin projections
of the two up quarks.}
\label{tabellea} 
\renewcommand{\arraystretch}{1.0}
\end{table}

\begin{table}
\renewcommand{\arraystretch}{1.3}
\begin{center}
\begin{tabular}{l|l|l|l}
&  
Lorentz-structure 
&  
Light-cone projection   
& 
nomenclature
\\ \hline
twist-3   
&  $ \left(C i \si_{\perp z} \right) \otimes \ga^\perp \!\not\!{z} $    
& $u^+_\up u^+_\up d^+_\down$
& $T_1(x_i)$ \\ \hline
twist-4
&  $ \left(C i \si_{\perp z} \right) \otimes \ga^\perp \!\not\!{p} $    
& $u^+_\down u^+_\down d^-_\down$
& $T_2(x_i)$ 
\\ \hline
& $ \left(C \!\not\!{z}\!\not\!{p}\right)  \otimes \!\not\!{z}$    
& $u^+_\up u^-_\up  d^+_\up$
& $- \left[T_3 -  T_7 - S_1 - P_1\right](x_i)$ 
\\ \hline
& $ \left(C \!\not\!{z}\!\not\!{p}\right)  \otimes \!\not\!{z}$    
& $u^+_\down u^-_\down  d^+_\up$
& $- \left[T_3+ T_7 - S_1 + P_1\right](x_i)$ 
\\ \hline
twist-5
&  $ \left(C i \si_{\perp p} \right) \otimes \ga^\perp \!\not\!{z} $    
& $u^-_\down u^-_\down d^+_\down$
& $T_5(x_i)$ 
\\ \hline
& $ \left(C \!\not\!{p}\!\not\!{z}\right)  \otimes \!\not\!{p}$    
& $u^-_\up u^+_\up  d^-_\up$
& $ -\left[T_4-  T_8 - S_2 - P_2\right](x_i)$ 
\\ \hline
& $ \left(C \!\not\!{p}\!\not\!{z}\right)  \otimes \!\not\!{p}$    
& $u^-_\down u^+_\down  d^-_\up$
& $ - \left[T_4+  T_8 - S_2 + P_2\right](x_i)$ 
\\ \hline
twist-6   
&  $ \left(C i \si_{\perp p} \right) \otimes \ga^\perp \!\not\!{p} $    
& $u^-_\up u^-_\up d^-_\down$
& $T_6(x_i)$ \\ \hline
\end{tabular}
\end{center} 
\caption[]{\sf 
 The remaining eight nucleon distribution amplitudes 
that enter the expansion in \Gl{defdisamp}.}
\label{tabelleb} 
\renewcommand{\arraystretch}{1.0}
\end{table}
In the following subsections we give the complete set of 
distribution amplitudes starting from twist-3 to twist-6.
The numerical values of the expansion parameters can be obtained
from Table \ref{tabelle3} in the text.
\subsection{Twist-3 distribution amplitudes}
\beqar{All-twist-3} 
V_1(x_i,\mu) &=& 120 x_1 x_2 x_3 \left[\phi_3^0(\mu) + 
\phi_3^+(\mu) (1- 3 x_3)\right]\,,
\nn \\
A_1(x_i,\mu) &=& 120 x_1 x_2 x_3 (x_2 - x_1) \phi_3^-(\mu) \,,
\nn \\
T_1(x_i,\mu) &=& 120 x_1 x_2 x_3 
\left[\phi_3^0(\mu) + \frac12\left(\phi_3^- - \phi_3^+\right)(\mu) 
(1- 3 x_3)\right]\,.
\eeqar

\subsection{Twist-4 distribution amplitudes}

\beqar{All-twist-4} 
V_2(x_i,\mu)  &=& 24 x_1 x_2 
\left[\phi_4^0(\mu)  + \phi_4^+(\mu)  (1- 5 x_3)\right] \,,
\nn\\
A_2(x_i,\mu)  &=& 24 x_1 x_2 (x_2 - x_1) \phi_4^-(\mu) \,,
\nn \\
V_3(x_i,\mu)  &=& 
12 x_3 \left[
\psi_4^0(\mu)(1-x_3)  + \psi_4^-(\mu) (x_1^2 + x_2^2 - x_3
(1-x_3) ) + \psi_4^+(\mu)( 1-x_3 - 10 x_1 x_2)\right]\,,
\nn \\
A_3(x_i,\mu) &=& 12 x_3 (x_2-x_1) 
\left[\left(\psi_4^0(\mu) + \psi_4^+(\mu)\right)
+ \psi_4^-(\mu) (1-2 x_3) \right] \,,
\nn \\
T_3(x_i,\mu)  &=& 
6 x_3 \left[
(\xi_4^0 + \phi_4^0 + \psi_4^0)(\mu)(1-x_3)  + 
(\xi_4^- + \phi_4^- - \psi_4^-)(\mu) (x_1^2 + x_2^2 - x_3 (1-x_3) ) 
\right. \nn \\ && \left.\qquad
+ (\xi_4^+ + \phi_4^+ + \psi_4^+)(\mu) ( 1-x_3 - 10 x_1 x_2) \right]\,,
\nn \\
T_7(x_i,\mu)  &=& 
6 x_3 \left[
(- \xi_4^0 + \phi_4^0 + \psi_4^0)(\mu)(1-x_3)  + 
(- \xi_4^- + \phi_4^- - \psi_4^-)(\mu) (x_1^2 + x_2^2 - x_3 (1-x_3) ) 
\right. \nn \\ && \left.\qquad
+ (- \xi_4^+ + \phi_4^+ + \psi_4^+)(\mu) ( 1-x_3 - 10 x_1 x_2) \right]\,,
\nn \\
T_2(x_i,\mu) 
&=& 24 x_1 x_2 \left[
\xi_4^0(\mu) + \xi_4^+(\mu)( 1- 5 x_3)\right]\,,
\nn \\
S_1(x_i,\mu) &=& 
6 x_3 (x_2-x_1) \left[
(\xi_4^0 + \phi_4^0 + \psi_4^0 + \xi_4^+ + \phi_4^+ + \psi_4^+)(\mu)   
+ (\xi_4^- + \phi_4^- - \psi_4^-)(\mu)(1-2 x_3) \right] \,,
\nn \\
P_1(x_i,\mu) &=& 
6 x_3 (x_2-x_1) \left[
(\xi_4^0 - \phi_4^0 - \psi_4^0 + \xi_4^+ - \phi_4^+ - \psi_4^+)(\mu)  
+ (\xi_4^- - \phi_4^- + \psi_4^-)(\mu)(1-2 x_3) \right] \,.
\nn \\
\eeqar

\subsection{Twist-5 distribution amplitudes}

\beqar{All-twist-5} 
V_5(x_i,\mu) &=& 6 x_3 
\left[\phi_5^0(\mu)  + \phi_5^+(\mu)(1- 2 x_3)\right]\,, 
\nn\\
A_5(x_i,\mu) &=& 6 x_3 (x_2-x_1) \phi_5^-(\mu) \,, 
\nn\\
V_4(x_i,\mu) &=& 3 \left[
\psi_5^0(\mu)(1-x_3)  + \psi_5^-(\mu)\left(2 x_1x_2 - x_3(1-x_3)\right)
+ \psi_5^+(\mu)(1-x_3 - 2 (x_1^2 +  x_2^2))\right] \,,
\nn\\
A_4(x_i,\mu) &=& 3 (x_2 -x_1)
\left[- \psi_5^0(\mu) + \psi_5^-(\mu) x_3 + \psi_5^+(\mu)(1- 2 x_3) \right]\,,
\nn \\
T_4(x_i,\mu) &=& \frac32 \left[
\left(\xi_5^0 + \psi_5^0 + \phi_5^0\right) (\mu)(1-x_3)  + 
\left(\xi_5^- + \phi_5^- - \psi_5^-\right)
(\mu)\left(2 x_1x_2 - x_3(1-x_3)\right)
\right. \nn \\ && \left.\qquad
+ \left(\xi_5^+ + \phi_5^+ + \psi_5^+\right)
(\mu)(1-x_3 - 2 (x_1^2 +  x_2^2))\right] \,,
\nn \\
T_8(x_i,\mu) &=& \frac32 \left[
\left(\psi_5^0 + \phi_5^0 - \xi_5^0\right)(\mu)(1-x_3)  + 
\left(\phi_5^- - \psi_5^- - \xi_5^-\right)(\mu)
\left(2 x_1x_2 - x_3(1-x_3)\right)
\right. \nn \\ && \left.\qquad
+ \left(\phi_5^+ + \psi_5^+ - \xi_5^+\right)(\mu)
(1-x_3 - 2 (x_1^2 +  x_2^2))\right] \,,
\nn \\
T_5(x_i,\mu) &=& 6 x_3 \left[ 
\xi_5^0(\mu) + \xi_5^+(\mu)( 1- 2 x_3)\right] \,,
\nn \\
S_2(x_i,\mu) &=& \frac32 (x_2 -x_1) 
\left[- \left(\psi_5^0 + \phi_5^0 + \xi_5^0\right)(\mu) + 
\left(\xi_5^- + \phi_5^- - \psi_5^0 \right)(\mu) x_3 
\right. \nn \\ && \left.\qquad\qquad
+ \left(\xi_5^+ + \phi_5^+ + \psi_5^0 \right)(\mu) (1- 2 x_3)\right]\,,
\nn \\
P_2(x_i,\mu) &=& \frac32 (x_2 -x_1)
\left[\left(\psi_5^0 + \psi_5^0 - \xi_5^0\right)(\mu) +
\left(\xi_5^- - \phi_5^- + \psi_5^0 \right)(\mu) x_3 
\right. \nn \\ && \left.\qquad\qquad
+ \left(\xi_5^+ - \phi_5^+ - \psi_5^0 \right)(\mu) (1- 2 x_3)\right]\,,
\eeqar

\subsection{Twist-6 distribution amplitudes}

\beqar{All-twist-6} 
V_6(x_i,\mu) &=& 2 \left[\phi_6^0(\mu) +  \phi_6^+(\mu) (1- 3 x_3)\right]\,,
\nn \\
A_6(x_i,\mu) &=& 2 (x_2 - x_1) \phi_6^- \,,
\nn \\
T_6(x_i,\mu) &=& 2 \left[\phi_6^0(\mu) + 
\frac12\left(\phi_6^--\phi_6^+\right) (1- 3 x_3)\right]\,. 
\eeqar

\vfill
\eject

\end{document}